\documentclass[12pt,twocolumn]{aastex63}

\usepackage{xcolor}
\usepackage{cancel}
\usepackage{physics}
\usepackage{multirow}

\submitjournal{ApJ 1 March 2022 \\ Accepted for publication 27 May 2022}
\shorttitle{AGILE observations of GRB~220101A}
\shortauthors{Ursi et al.}

\begin{document}

\title{AGILE Observations of GRB~220101A\\A ``New Year's Burst'' with an Exceptionally Huge Energy Release}

\correspondingauthor{Alessandro Ursi}
\email{alessandro.ursi@inaf.it}

\author[0000-0002-7253-9721]{A. Ursi}
\affiliation{\scriptsize INAF/IAPS, via del Fosso del Cavaliere 100, I-00133 Roma (RM), Italy}

\author[0000-0002-4724-7153]{M.~Romani}
\affiliation{\scriptsize Osservatorio Astronomico di Brera, Via Brera 28, I-20121 Milano (MI), Italy}

\author[0000-0002-9332-5319]{G.~Piano}
\affiliation{\scriptsize INAF/IAPS, via del Fosso del Cavaliere 100, I-00133 Roma (RM), Italy}

\author[0000-0003-3455-5082]{F.~Verrecchia}
\affiliation{\scriptsize SSDC/ASI, via del Politecnico snc, I-00133 Roma (RM), Italy}
\affiliation{\scriptsize INAF/OAR, via Frascati 33, I-00078 Monte Porzio Catone (RM), Italy}

\author[0000-0003-2501-2270]{F.~Longo}
\affiliation{\scriptsize Dipartimento di Fisica, Universit\`a di Trieste and INFN, via Valerio 2, I-34127 Trieste (TR), Italy}

\author[0000-0001-6661-9779]{C.~Pittori}
\affiliation{\scriptsize SSDC/ASI, via del Politecnico snc, I-00133 Roma (RM), Italy}
\affiliation{\scriptsize INAF/OAR, via Frascati 33, I-00078 Monte Porzio Catone (RM), Italy}

\author[0000-0003-2893-1459]{M.~Tavani}
\affiliation{\scriptsize INAF/IAPS, via del Fosso del Cavaliere 100, I-00133 Roma (RM), Italy}
\affiliation{\scriptsize Universit\`a degli Studi di Roma Tor Vergata, via della Ricerca Scientifica 1, I-00133 Roma (RM), Italy}







\author[0000-0001-6347-0649]{A.~Bulgarelli}
\affiliation{\scriptsize INAF/OAS, via Gobetti 101, I-40129 Bologna (BO), Italy}

\author[0000-0001-8877-3996]{M.~Cardillo}
\affiliation{\scriptsize INAF/IAPS, via del Fosso del Cavaliere 100, I-00133 Roma (RM), Italy}

\author[0000-0001-8100-0579]{C.~Casentini}
\affiliation{\scriptsize INAF/IAPS, via del Fosso del Cavaliere 100, I-00133 Roma (RM), Italy}
\affiliation{\scriptsize INFN Sezione di Roma 2, via della Ricerca Scientifica 1, I-00133 Roma (RM), Italy}

\author[0000-0001-6877-6882]{P.~W.~Cattaneo}
\affiliation{\scriptsize INFN Sezione di Pavia, via U. Bassi 6, I-27100 Pavia (PV), Italy}


\author[0000-0003-4925-8523]{E.~Costa}
\affiliation{\scriptsize INAF/IAPS, via del Fosso del Cavaliere 100, I-00133 Roma (RM), Italy}




\author[0000-0002-7617-3421]{M.~Feroci}
\affiliation{\scriptsize INAF/IAPS, via del Fosso del Cavaliere 100, I-00133 Roma (RM), Italy}


\author[0000-0002-6082-5384]{V.~Fioretti}
\affiliation{\scriptsize INAF/OAS, via Gobetti 101, I-40129 Bologna (BO), Italy}

\author[0000-0002-0709-9707]{L.~Foffano}
\affiliation{\scriptsize INAF/IAPS, via del Fosso del Cavaliere 100, I-00133 Roma (RM), Italy}







\author[0000-0002-6311-764X]{F.~Lucarelli}
\affiliation{\scriptsize SSDC/ASI, via del Politecnico snc, I-00133 Roma (RM), Italy}
\affiliation{\scriptsize INAF/OAR, via Frascati 33, I-00078 Monte Porzio Catone (RM), Italy}

\author[0000-0002-4000-3789]{M.~Marisaldi}
\affiliation {\scriptsize Birkeland Centre for Space Science, Department of Physics and Technology, University of Bergen, Norway}
\affiliation{\scriptsize INAF/OAS, via Gobetti 101, I-40129 Bologna (BO), Italy}


\author[0000-0002-7704-9553]{A.~Morselli}
\affiliation{\scriptsize INFN Sezione di Roma 2, via della Ricerca Scientifica 1, I-00133 Roma (RM), Italy}

\author[0000-0001-6897-5996]{L.~Pacciani}
\affiliation{\scriptsize INAF/IAPS, via del Fosso del Cavaliere 100, I-00133 Roma (RM), Italy}


\author[0000-0002-4535-5329]{N.~Parmiggiani}
\affiliation{\scriptsize INAF/OAS, via Gobetti 101, I-40129 Bologna (BO), Italy}



\author{P.~Tempesta}
\affiliation{\scriptsize Telespazio SpA, Centro Spaziale del Fucino 23, Piana del Fucino, Via Cintarella, I-67050, Ortucchio (AQ), Italy}

\author[0000-0002-3180-6002]{A.~Trois}
\affiliation{\scriptsize INAF - Osservatorio Astronomico di Cagliari, via della Scienza 5, I-09047 Selargius (CA), Italy}

\author[0000-0003-1163-1396]{S.~Vercellone}
\affiliation{\scriptsize INAF - Osservatorio Astronomico di Brera, via E. Bianchi 46, I-23807 Merate (LC), Italy}



\begin{abstract}

We report the AGILE observations of GRB~220101A, which took place at the beginning of 1st January 2022 and was recognized as one of the most energetic gamma-ray bursts (GRBs) ever detected since their discovery. The AGILE satellite acquired interesting data concerning the prompt phase of this burst, providing an overall temporal and spectral description of the event in a wide energy range, from tens of keV to tens of MeV. Dividing the prompt emission into three main intervals, we notice an interesting spectral evolution, featuring a notable hardening of the spectrum in the central part of the burst. The average fluxes encountered in the different time intervals are relatively moderate, with respect to those of other remarkable bursts, and the overall fluence exhibits a quite ordinary value among the GRBs detected by MCAL. However, GRB~220101A is the second farthest event detected by AGILE, and the burst with the highest isotropic equivalent energy of the whole MCAL GRB sample, releasing $E_{iso}=2.54\cdot10^{54}$~erg and exhibiting an isotropic luminosity of $L_{iso}=2.34\cdot10^{52}$~erg~s$^{-1}$ (both in the 400~keV--10~MeV energy range).

We also analyzed the first $10^6$~s of the afterglow phase, using the publicly available Swift XRT data, carrying out a theoretical analysis of the afterglow, based on the forward shock model. We notice that GRB~220101A is with high probability surrounded with a wind-like density medium, and that the energy carried by the initial shock shall be a fraction of the total $E_{iso}$, presumably near $\sim50\%$. 

\end{abstract}



\section{Introduction}

Gamma-ray bursts (GRBs) are powerful transient gamma-ray emissions, releasing isotropic equivalent energies on the order of $E_{iso}\gtrsim10^{52}$~erg, and representing the most luminous events observed in the universe to date \citep{Gehrels&Meszaros2012}. Discovered in the late 1960s \citep{Klebesadel1973}, these bursts are produced by ultra-relativistic particles, accelerated in extra-galactic engines. Depending on their energy spectrum and their $T_{90}$ duration \citep[i.e., the time interval over which the central $90\%$ of their cumulative counts above the background is detected,][]{Kouveliotou1993}, GRBs are conventionally divided into short GRBs ($T_{90}<2$~s) and long GRBs ($T_{90}>2$~s). Short GRB emission can extend above MeV energies during the prompt phase \citep[e.g., GRB~090510][]{Ackermann2010,Depasquale2010,Giuliani2010} and are associated with the merger of two compact extremely massive objects, such as neutron star-neutron star (BNS), or black hole-neutron star (NSBH) systems \citep{Goldstein2017,Connaughton2017}. Long GRBs have softer spectra and are associated with type Ic core-collapse supernovae, possibly occurring in the presence of an evolved star companion (e.g., a neutron star, or a black hole) \citep{Vedrenne2009}.

GRBs release most of their energy during the prompt phase, in the few keV -- few MeV energy range. Their spectrum can be usually described by means of a Band model, a smoothly joint broken power-law with a well-defined peak energy \citep{Band1993}. Some bursts can show extra high-energy components, which extend the spectrum up to hundreds of MeV -- GeV energies, requiring additional power-laws, or cutoff power-laws, in the spectral model \citep[e.g., GRB~910503, GRB~930131,GRB~941017,GRB~080514B,GRB~131108A][]{Schneid1992,Sommer1994,Gonzalez2003,Giuliani2008,Giuliani2014}. These high-energy components can emerge either during the prompt phase, probably produced internally due to inverse Compton-scattered synchrotron photons of the prompt \citep{Bosnjak2009}, or during the early afterglow phases, coming as delayed emissions, probably arising from external shocks traveling in the circumburst medium \citep{Ackermann2013}.

\subsection{GRB~220101A}

GRB~220101A is a long GRB occurred at 2022-01-01 05:11:13 (UT), first detected by Swift BAT \citep{Tohuvavohu2022,Markwardt2022}, and successively revealed in the low to intermediate energy range by Swift XRT \citep{Osborne2022,Dai2022}, Fermi LAT \citep{Arimoto2022}, Swift UVOT \citep{Kuin2022}, AGILE \citep{Ursi2022gcn}, Fermi GBM \citep{Lesage2022}, and Konus-Wind \citep{Tsvetkova2022a,Tsvetkova2022b}. The burst was promptly localized by imaging detectors onboard Swift and Fermi at coordinates RA,Dec=$0.09^{\circ},31.76^{\circ}$. In addition to X/gamma-ray detections, a number of fast-response optical observations of the GRB have been carried out, which allowed to reveal an associated optical transient with redshift z=4.62 \citep{Fu2022,Perley2022,Fynbo2022,Tomasella2022}. Such intermediate value of $z$ led to a preliminary estimate of the associated isotropic equivalent energy equal to $E_{iso}\sim3.6\cdot10^{54}$~erg, making GRB~220101A one of the most energetic events in the history of GRBs \citep{Atteia2022,Ruffini2022b}. Depending on the energy range, the event duration was reported up to $\sim280$~s, with a prompt phase exhibiting a bright and complex multi-peaked time profile, consisting of different episodes.

In this paper, we focus on the properties of this remarkable GRB, which is the event with the highest associated $E_{iso}$ detected by AGILE, to date. Such huge reconstructed isotropic energy release is largely ascribed to the huge luminosity distance of $\sim46,483$~Mpc associated to its estimated redshift. We analyze the AGILE scientific ratemeters data, which cover about 4 orders of magnitude in energy, to reconstruct the GRB time profile and to study its spectral evolution. We also analyzed the Swift XRT data concerning the afterglow emission, to model the progenitor circumburst density profile and characterize the scenario in which the shock propagated.

Our results are in general agreement with the preliminary results reported in various GCN circulars delivered by Swift, Fermi, and Konus-Wind, and provide information on the temporal and spectral properties of GRB~220101A.

\section{The AGILE satellite}
\label{TheAGILEsatellite}

AGILE (Astrorivelatore Gamma ad Immagini LEggero) is a satellite of the Italian Space Agency (ASI), launched in 2007 and devoted to high-energy astrophysics \citep{Tavani2008b}. It consists of an imaging gamma-ray silicon tracker (30~MeV--50~GeV), a coded mask X-ray imager SuperAGILE (SA, 18--60~keV), an all sky Mini-CALorimeter (MCAL, 0.4--100~MeV), and an Anti-Coincidence system (AC, 50--200~keV). The silicon tracker, MCAL, and AC detectors form the so-called Gamma-Ray Imaging Detector (GRID) \citep{Barbiellini2001b, Prest2003}. The AGILE satellite orbits at $\sim500$~km altitude, in a low-earth, quasi-equatorial orbit. Due to a failure of its reaction wheel occurred in 2009, AGILE currently spins about its sun-pointing axis, with a frequency of $\sim7$~min$^{-1}$, monitoring about $80\%$ of the available sky with its imaging detectors. The AGILE data are downloaded to ground at each passage over the ASI Ground Station in Malindi, Kenya, and then processed at the AGILE data center in ASI-SSDC \citep{Pittori2019}, delivering scientific alerts for high-energy transients within 20~minutes -- 2~hours from the onboard acquisition. In this work, we carry out analysis of the AGILE MCAL and AGILE scientific ratemeters data, illustrated in detail in the following sections.

\subsection{The AGILE MCAL}
\label{MCALlogic}

The Mini-CALorimeter (MCAL) is a non-imaging, all-sky scintillation detector \citep{Labanti2009,Marisaldi2008}. It consists of 30 CsI(Tl) bars, providing a total on-axis geometrical area of 1400~cm$^2$ and an effective area of $\sim300$~cm$^2$ at 1~MeV. MCAL is a triggered instrument, whose trigger logic works on different timescales (0.293~ms, 1~ms, 16~ms, 64~ms, 256~ms, 1024~ms, and 8192~ms): this allows to detect both short-duration and long-duration high-energy transients, such as GRBs \citep{Galli2013,Ursi2022catalog}, or terrestrial gamma-ray flashes \citep{Marisaldi2014,Maiorana2020,Lindanger2020}, and to carry out searches for possible gamma-ray signatures associated to other astrophysical events, such as gravitational wave events \citep{Ursi2019,Verrecchia2017a,Ursi2022gw}. Given its energy range, MCAL is mostly focused on the detection of hard-spectrum bursts: in particular, between 2007 and 2020, the AGILE MCAL detected more than 500~GRBs, most of which exhibited a non-negligible spectral component above 1~MeV \citep{Ursi2022catalog}\footnote{https://www.ssdc.asi.it/mcal2grbcat/}.

The AGILE team has developed automatic pipelines to perform quick analysis of MCAL data, once the satellite telemetry is downloaded at the Malindi ground station \citep{Parmiggiani2021}. An offline algorithm carries out blind searches for impulsive transients and deliver prompt communication to the AGILE team, whenever a gamma-ray transient is identified. Moreover, in case of a GRB detection, an automatic Gamma-ray Coordinates Network (GCN) notice is promptly delivered to the scientific community\footnote{https://gcn.gsfc.nasa.gov/agile\_mcal.html}. The AGILE pipelines also perform fast follow-up of external alerts from other space missions or facilities (e.g., Swift, IceCube, LIGO-Virgo), allowing a prompt reaction of the AGILE team in the multi-wavelength/multi-messenger context \citep{Bulgarelli2019a,Bulgarelli2019b}.

\subsection{The AGILE scientific ratemeters}

Data acquired by the GRID, SA, MCAL, and AC detectors are continuously stored in telemetry, with a time resolution of 0.512~s (for SA) and 1.024~s (for GRID, MCAL, and AC). Such data streams, called scientific ratemeter (RM) data, are independent on any onboard trigger and provide a continuous monitoring of the X/gamma-ray background in all the detectors. Although aimed at monitoring the background variation through the orbital phases, the AGILE RM data clearly reveal high-energy transients, such as GRBs \citep{Ursi2022catalog}, soft gamma repeaters (SGRs) \citep[e.g.,][]{Tavani2021}, and solar flares \citep[e.g.,][]{Ursi2020solar}, and can serve as independent detectors as well. In particular, the SA and AC RMs, sensitive in the hard X-ray energy range, detected more than 700~GRBs between 2007 and 2020, whereas the MCAL RMs detected more than 500~events. Despite the coarse time resolution, the MCAL RM data are acquired in 11 energy channels, covering an energy range from $\sim200$~keV to more than $\sim180$~MeV, which allow to reconstruct a preliminary energy spectrum of the detected events. In particular, the joint usage of SuperAGILE and MCAL RM data allow to have 12 energy spectral channels, whose energy ranges are: CH0 [18--60~keV], CH1 [175--350~keV], CH2 [350--700~keV], CH3 [0.7--1.4~MeV], CH4 [1.4--2.8~MeV], CH5 [2.8--5.6~MeV], CH6 [5.6--11.2~MeV], CH7 [11.2-22.4~MeV], CH8 [22.4--44.8~MeV], CH9 [44.8--89.6~MeV], CH10 [89.6--179.2~MeV], CH11 [$>$179.2~MeV]. The AC data are not calibrated in energy and are therefore not used for spectral analysis. The last channel CH11 is affected by large errors on the reconstructed count rate and it is therefore rejected. MCAL CH1 is considered only for the MCAL RM data, that are integrated onboard with a fixed 1.024~s time resolution; on the other hand, when dealing with MCAL triggered data, which are typically rebinned and analyzed on millisecond - tens of millisecond timescales, we only consider the 0.4-100~MeV energy range, that offers a more reliable reconstructed energetics. The AGILE RM data are routinely calibrated, comparing the detected GRBs and SGR bursts with the fluxes and spectra reported for the same events by other space missions.


\section{The prompt phase of GRB~220101A}
\label{AGILEobservationsofGRB220101A}

The AGILE satellite detected the GRB~220101A at $T_0$ = 2022-01-01 05:11:13 UT (hereafter called $T_0$). The prompt phase of the event was clearly visible in the SA, AC, and MCAL scientific RM data. In addition, the event triggered an MCAL photon-by-photon onboard data acquisition, with 4~$\mu$s time resolution, which lasted $\sim34~s$ and covered the time interval from $T_0+8.43$~s and $T_0+42.71$~s. In particular, the triggered logic was the 64~ms MCAL timescale.

GRB~220101A took place during the commissioning phase of the Imaging X-ray Polarimetry Explorer (IXPE) space mission, launched 9 December 2021, which employed a large amount of available telemetry at the ASI ground station in Malindi. As a consequence, the number of served passages to download the AGILE data was largely reduced. The AGILE satellite was therefore operating in a non-optimized, reduced onboard configuration, designed to save onboard mass-memory throughout more consequent orbits. This translated into the non-availability of SA and GRID imaging data, and into a non-optimized onboard MCAL trigger logic, which prevented the full acquisition of the GRB with high time resolution. In this perspective, given the relatively large time duration of the event, the AGILE RM data are fundamental to provide an overall picture of the burst, both for the study of the light-curve time profile and spectral properties. Moreover, the partial MCAL data acquisition allows to focus on the transition between a first softer stage of the burst prompt and the onset of a larger flux, higher-energy emission.

\subsection{Satellite attitude}
\label{satelliteattitude}

\begin{figure}
\centering
  \includegraphics[width=1.0\linewidth]{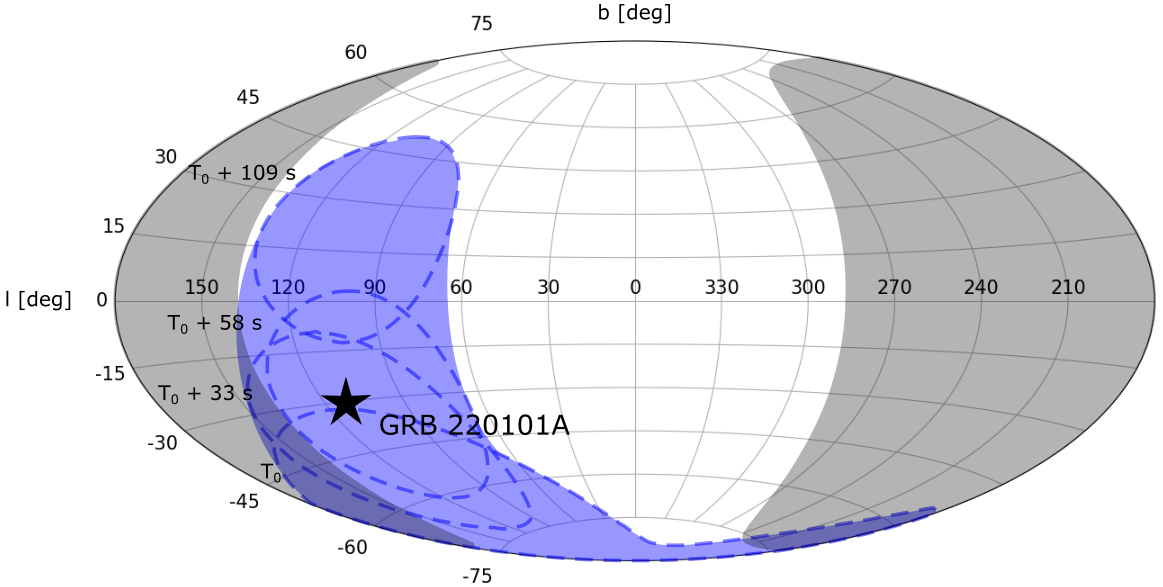}
\caption{SuperAGILE $40^{\circ}$ field of view (blue shaded region) in galactic coordinates, in the time interval from $T_0$ to $T_0+109$~s. The dashed lines represent the contours at the transitions between subsequent time intervals \textit{A}, \textit{B}, and \textit{C} (reported on the left side). The localization region of GRB~220101A ($l=111.79^{\circ}$,$b=-30.14^{\circ}$, indicated by the star) lays within the field of view for most of the prompt phase, with the best on-axis configuration ($\sim0.7^{\circ}$) reached at $T_0+49$~s.}
\label{FOV}
\end{figure}

As the AGILE satellite spins about its sun-pointing axis and as GRB~220101A exhibits a rather long duration, it is important to evaluate how the AGILE boresight changed during the occurrence of the burst. This evaluation is fundamental for what concerns the SA detector, whose detection capabilities are reliable only if the source falls inside its field of view. Also, it allows to retrieve the corresponding MCAL response matrices needed to perform spectral analysis. Fig.~\ref{FOV} reports the SA $40^{\circ}$ field of view between $T_0$ and $T_0+109$~s: it can be noticed that the GRB~220101A localization region was inside the field of view for most of its duration, reaching the most on-axis configuration ($0.7^{\circ}$) at $T_0+49$~s. As a consequence, for times $<T_0-12$~s and $>T_0+135$~s, we do not consider the SA data, as the source was observed under a too large off-axis angle, preventing any reliable reconstruction of the count rate and flux.

\subsection{Temporal characteristics}

\begin{figure*}
\centering
  \includegraphics[width=1.0\linewidth]{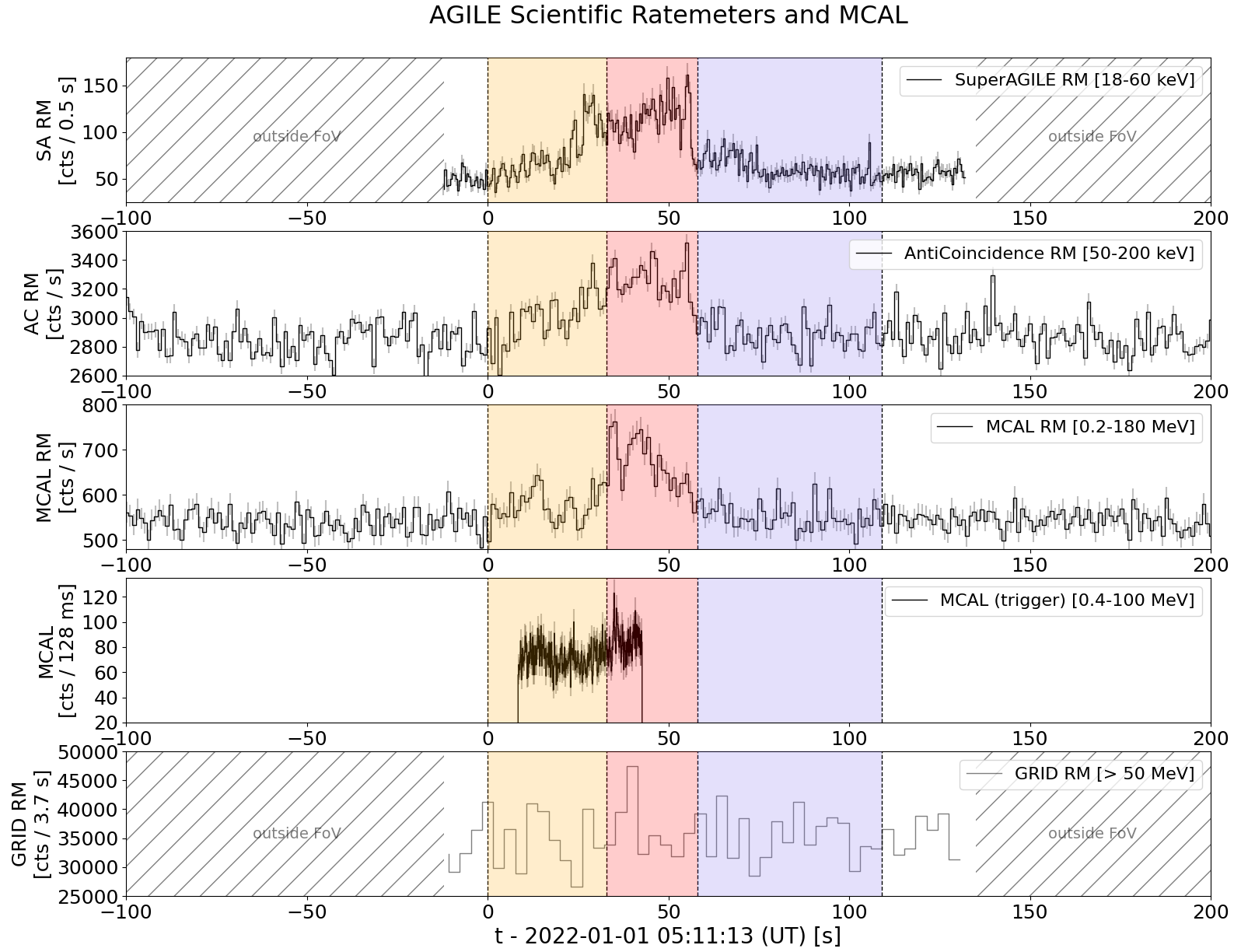}
\caption{Light-curves of the prompt phase of GRB~220101A, acquired by AGILE (and associated $\pm1\sigma$ error bars): the first three panels display the SuperAGILE, Anti-Coincidence, and MCAL scientific ratemeters (RM) data, acquired in the related energy ranges. The fourth panel shows the partial MCAL high time resolution trigger issued onboard, covering the time interval between $T_0+8.43$~s and $T_0+42.71$~s. Finally, the bottom panel reports the GRID RM data in the highest-energy regime, integrated on the whole available sky inside the detector field of view. We excluded the SuperAGILE and GRID data acquired before $T_0-12$~s and after $T_0+135$~s, as the source localization region was outside the detectors' field of view. We divide the light-curves into three intervals: \textit{A} (from $T_0$ to $T_0+33$~s, yellow region), \textit{B} (from $T_0+33$~s to $T_0+58$~s, red region), and \textit{C} (from $T_0+58~s$ to $T_0+109$~s, violet region). The corresponding parts of intervals \textit{A} and \textit{B} covered by the MCAL trigger acquisition are referred to as \textit{interval-a1} and \textit{interval-b1}.}
\label{RM_LC}
\end{figure*}

\begin{table}
\centering
\caption{Properties of the time intervals of GRB~220101A.}
\label{tabintervals}
\begin{center}
\begin{tabular}{|c|c|c|c|}
\hline
\multicolumn{4}{|c|}{Scientific Ratemeters data} \\ \hline
interval    & duration & time start & time stop         \\ \hline
A           & 33.0~s     & $T_0$ & $T_0+33.0$~s     \\ \hline
B           & 25.0~s     & $T_0+33.0$~s & $T_0+58.0$~s  \\ \hline
C           & 51.0~s     & $T_0+58.0$~s & $T_0+109.0$~s \\ \hline \hline
\multicolumn{4}{|c|}{MCAL triggered data acquisition} \\ \hline
interval    & duration & time start & time stop  \\ \hline
a1          & 24.6~s     & $T_0+8.4$~s & $T_0+33.0$~s   \\ \hline
b1          & 9.7~s     & $T_0+33.0$~s & $T_0+42.7$~s  \\ \hline
\end{tabular}
\end{center}
Upper block: Details of the three time intervals A, B, and C of GRB~220101A, as detected by the AGILE scientific ratemeters. Bottom block: Details of the two time intervals \textit{a1} and \textit{b1}, covered by the partial AGILE MCAL triggered data acquisition.
\end{table}

The temporal profile of GRB~220101A varies with respect to energy, from $\sim20$~keV to $\sim50$~MeV, with the bulk emission occurring during the central time interval. The overall light-curve consists of an initial slow flux enhancement, mostly dominated by X-ray/soft gamma-ray emission, followed by a central brigher multipeaked emission lasting $\sim25$~s and characterized by the peak emission of the MeV range. Finally, a dim broad emission lasting more than $\sim50$~s, fades to undetectable level and sets off the end of the prompt phase. The first three panels of Fig.~\ref{RM_LC} show the SA, AC Top, and MCAL RM light-curves, centered at $T_0$, with 0.512~s (for SA) and 1.024~s (for AC and MCAL) time resolution. The light-curves show a rather long and complex multi-peaked profile. As MCAL is the detector offering the best energy coverage for the burst spectrum, we divide the light-curve into three time intervals, on the basis of the GRB shape in the MCAL energy range: \textit{interval-A} (from $T_0$ to $T_0+33$~s, displayed in yellow), \textit{interval-B} (from $T_0+33$~s to $T_0+58$~s, displayed in red), and \textit{interval-C} (from $T_0+58~s$ to $T_0+109$~s, displayed in violet), reported in detail in Tab.~\ref{tabintervals}. As pointed out in Subsection~\ref{satelliteattitude}, we excluded all SA data acquired before $T_0-12$~s and after $T_0+135$~s, as the source localization region was outside the detector field of view. The fourth panel in figure reports the MCAL triggered acquisition, which covers a part of \textit{interval-A} and a part of \textit{interval-B}, hereby called \textit{interval-a1} and \textit{interval-b1}. Although partial, this acquisition allows to focus with better quality data on the transition between \textit{interval-A} and \textit{interval-B}, when the flux increases of more than a factor 2, especially in the MeV -- tens of MeV range. For completeness, the bottom panel of Fig.~\ref{RM_LC} reports also the GRID RM data. As illustrated in Section~\ref{TheAGILEsatellite}, the silicon tracker detector was not operational at the time GRB~220101A took place: as a consequence, the only available GRID data are those of the associated RMs, evaluated on the whole available portion of the sky inside the detector field of view (i.e., $\sim50^{\circ}$ with respect to the AGILE boresight). Similarly to SA, we excluded data acquired before $T_0-12$~s and after $T_0+135$~s, as the source was outside the tracker field of view, and the corresponding fluxes could not be properly reconstructed. Although these data refer to a large portion of the sky and are mostly populated by background noise, a rebinning of the GRID RM light-curve with $3.720$~s time resolution allows to identify a possible emission at $T_0+40$~s, when the source was only $\sim5^{\circ}$ off-axis. This event has a signal-to-noise ratio equal to 1.4 and a false alarm rate equal to $2.0\cdot10^{-4}$~Hz (considering 5.5~hours of data): we therefore point out the possibility that this spike could represent the signature of a high-energy emission (i.e., $>50$~MeV) taking place during \textit{interval-B}.

The initial emission episodes of GRB~220101A, occurring at about $T_0-50$~s, and reported in the Swift BAT  \citep{Tohuvavohu2022}, Fermi GBM \citep{Lesage2022}, and Konus-Wind \citep{Tsvetkova2022a} light-curves, have not been detected by AGILE, as for those times the source was outside the SA detector field of view. Nevertheless, since any significant emission was detected neither in the MCAL light-curve (which is not affected by field of view issues), we can infer that such initial episodes shall be dominated by X-ray emission. Similarly, no X-ray reflaring episodes are detected after $T_0+109$~s, as the source was outside the SA field of view.



GRB~220101A exhibited different durations, depending on the energy range. Adopting the algorithm illustrated in \citep{Koshut1996}, we calculated the corresponding $T_{50}$ and $T_{90}$ time durations, as seen by AGILE (i.e., the time over which the central $50\%$ or $90\%$ of the fluence is received, respectively). In particular, the event lasted $T_{50}=19.0\pm0.5$~s and $T_{90}=42.0\pm0.5$~s in the SA range, $T_{50}=18.5\pm1.0$~s and $T_{90}=44.5\pm1.0$~s and in the AC range, and $T_{50}=16.5\pm1.0$~s and $T_{90}=65.5\pm1.0$~s in the MCAL range.

\subsection{Spectral analysis}

\begin{figure}
\centering
\includegraphics[width=1.0\linewidth]{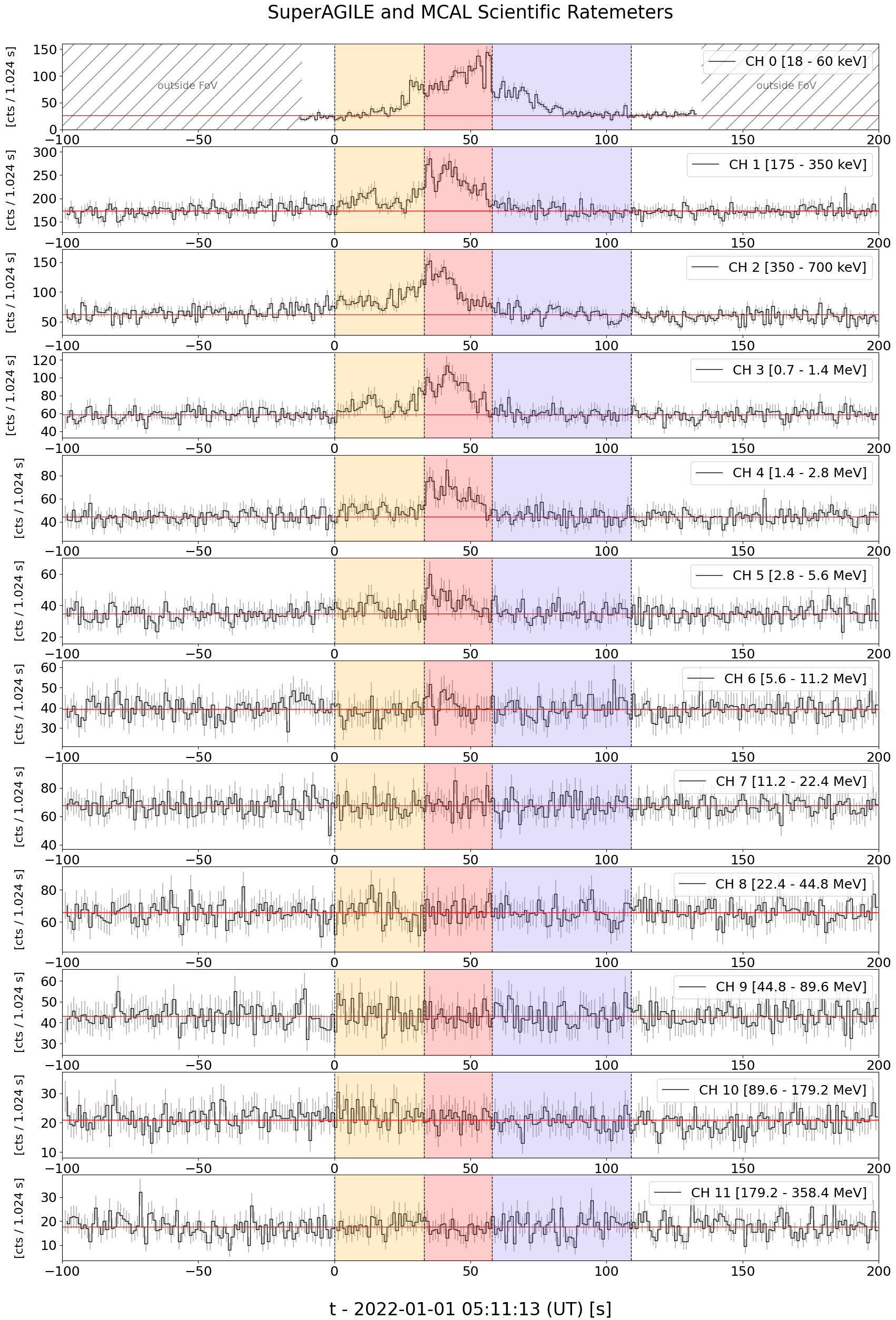}
\caption{SuperAGILE (first panel, CH0) and MCAL (other panels, CHs1-11) ratemeter light-curves of the prompt phase of GRB~220101A (and associated $\pm1\sigma$ error bars), with 1.024~s time resolution, and divided into 12 spectral energy channels. The SuperAGILE light-curve is rescaled on the basis of the associated effective area. The red line indicates the average background rate, estimated immediately before and after the reported time interval. Dashed lines divide the light-curve into \textit{A}, \textit{B}, and \textit{C} time intervals. It can be noticed a significant contribution up to CH7 (11.2--22.4~MeV), with a higher flux in \textit{interval-B}.}
\label{ALL_RM_LC}
\end{figure}

We performed spectral analysis of the prompt phase of GRB~220101A, taking advantage of both the AGILE RM data and the partial MCAL triggered data acquisition. Fig.~\ref{ALL_RM_LC} shows the GRB~220101A light-curve in the 12 spectral channels provided by both the SuperAGILE and MCAL RM data. In this plot, we rebinned the SuperAGILE light-curve to 1.024~s, to be consistent with the MCAL bin-width, and rescaled it on the basis of the associated effective area, which significantly changes throughout the burst duration. It can be noticed a significant contribution up to CH7 (11.2-22.4~MeV), especially in \textit{interval-B}, where the flux reaches its maximum in the MCAL energy range. For the spectral analysis, we only considered the first eight channels of the MCAL RM data, as the background-subtracted fluxes reconstructed above $\sim50$~MeV are affected by very low statistics and are not reliable.

\begin{figure*}
\centering
\includegraphics[width=0.9\linewidth]{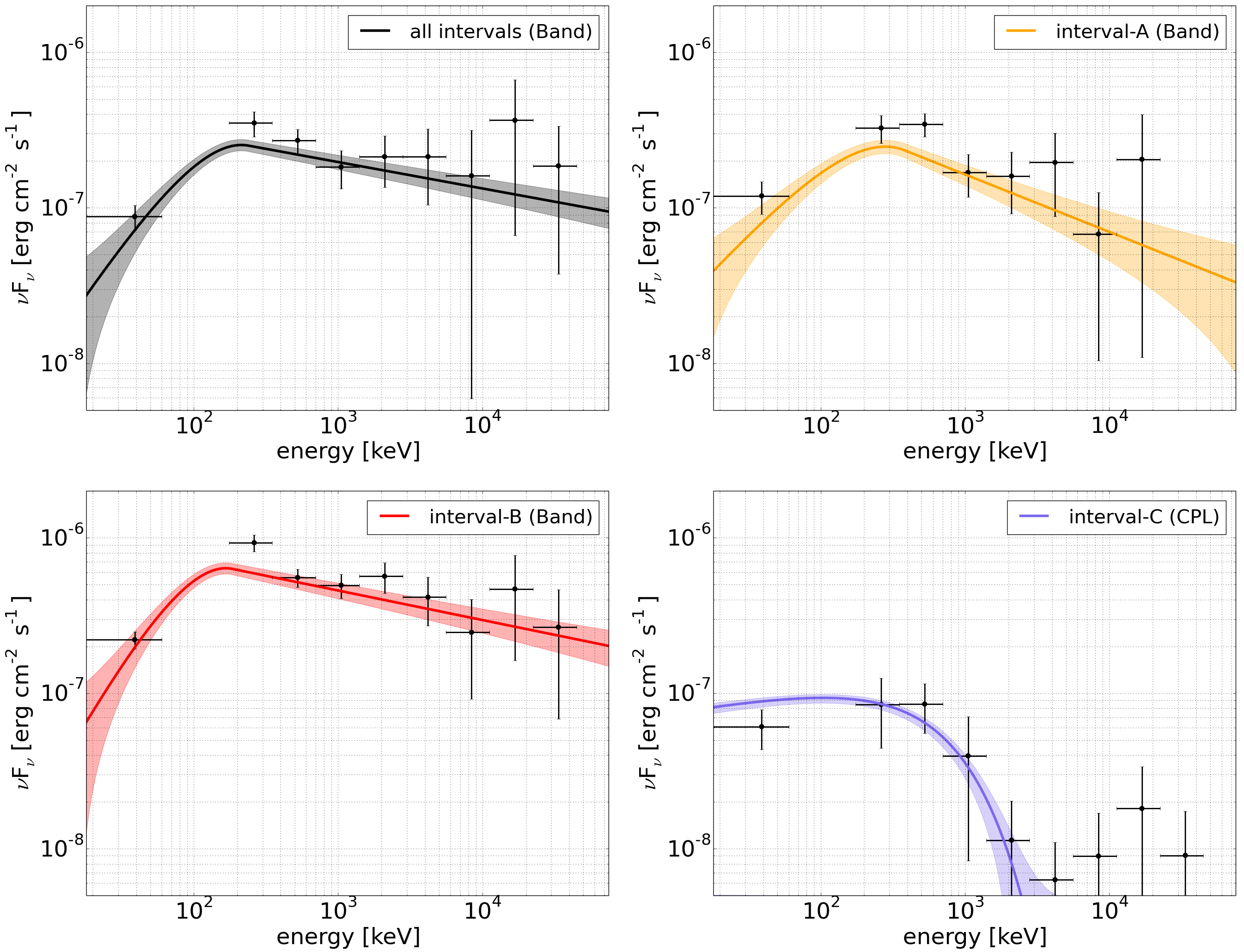}
\caption{Energy spectra of GRB~220101A, obtained from the SuperAGILE and MCAL ratemeters data. The spectra are time-integrated on the overall duration of the event (first panel, grey color), and in the subsequent different time intervals \textit{A} (yellow), \textit{B} (red), and \textit{C} (violet). The best-fit models and related $1\sigma$ confidence regions are illustrated with lines and shaded areas, and reported in the legend.}
\label{SPECTRA}
\end{figure*}

\begin{table*}
\centering
\caption{AGILE ratemeters and AGILE MCAL spectral analysis of GRB~220101A.}
\label{tabspectra}
\begin{center}
\begin{tabular}{|c|c|c|c|c|c|c|c|c|c|}
\hline
\multirow{2}{*}{interv} & \multirow{2}{*}{$\Delta t$ [s]} & \multicolumn{3}{|c|}{Band}                                        & \multicolumn{2}{|c|}{CPL}                              & PL                      & \multirow{2}{*}{$\chi^{2}_{red}$(dof)} & fluence [erg~cm$^{-2}$] \\ \cline{3-8}
                      &                    & $\alpha$                & $\beta$                 & $E_p$ [keV]                  & ph.ind.                 & $E_p$ [keV]                  & ph.ind.                & ~                                      & (18~keV--10~MeV)                         \\ \hline \hline
ABC                   & 109.0              & $-0.57_{-0.62}^{+1.38}$ & $-2.17_{-0.35}^{+0.21}$ & $213.89_{-184.41}^{+543.66}$ & --                      & --                           & --                     & 0.39 (4)                               & $(1.36\pm0.14)\cdot10^{-4}$             \\ \hline \hline
A                     &  33.0              & $-0.98_{-0.38}^{+0.39}$ & $-2.37_{-0.27}^{+0.13}$ & $279.68_{ -79.08}^{+462.32}$ & --                      & --                           & --                     & 0.40 (4)                               & $(3.93\pm0.39)\cdot10^{-5}$             \\ \hline
\multirow{2}{*}{B} & \multirow{2}{*}{51.0} & $-0.29_{-0.61}^{+0.43}$ & $-2.19_{-0.14}^{+0.10}$ & $166.81_{-143.35}^{+288.17}$ & --                      & --                           & --                     & 0.42 (7)                               & $(7.68\pm0.77)\cdot10^{-5}$             \\ \cline{3-10}
                      &                    & $-0.79_{-6.65}^{+4.05}$ & $-2.26_{-0.73}^{+0.07}$ & $151.49_{-135.63}^{+978.64}$ & --                      & --                           & $1.35_{-0.30}^{+1.29}$ & 0.20 (6)                               & $(8.05\pm0.80)\cdot10^{-5}$             \\ \hline
\multirow{2}{*}{C} & \multirow{2}{*}{51.0} & --                      & --                      & --                           & $1.85_{-2.15}^{+0.77}$  & $688.30_{-456.39}^{+692.42}$ & --                     & 1.01 (3)                               & $(7.67\pm0.77)\cdot10^{-6}$             \\ \cline{3-10}
                      &                    & --                      & --                      & --                           & --                      & --                           & $2.81_{-0.54}^{+0.84}$ & 0.15 (4)                               & $(4.01\pm0.40)\cdot10^{-6}$             \\ \hline \hline
a1                    &  24.6              & --                      & --                      & --                           & --                      & --                           & $2.51_{-0.54}^{+0.84}$ & 0.96 (86)                              & $(7.60\pm0.76)\cdot10^{-6}$             \\ \hline
b1                    &   9.7              & --                      & --                      & --                           & --                      & --                           & $2.25_{-0.21}^{+0.26}$ & 1.05 (86)                              & $(9.55\pm0.96)\cdot10^{-6}$             \\ \hline
\end{tabular}
\end{center}
Best fits of GRB~220101A obtained for intervals \textit{A}, \textit{B}, and \textit{C} from the SuperAGILE and MCAL ratemeters data, and for intervals \textit{a1} and \textit{b1} from MCAL data. For each interval, we report the related spectral parameters, reduced chi-square $\chi^2_{red}$ and number of degrees of freedom (dof), as well as fluences evaluated in the 18~keV--10~MeV energy range, and integrated in the corresponding time interval. Adopted models involve Band functions, power-laws (PLs), or cutoff power-laws (CPLs).
\end{table*}


Despite the coarse time resolution of the RM light-curves and the limited number of available spectral channels, the AGILE data can be used to provide a preliminary picture of the overall behavior of GRB~220101A and to point out some of the most salient points of its spectral evolution. The spectral fits were carried out using the XSPEC software package (version 12.12.0) \citep{Arnaud1996}. For each interval, we tested more spectral models, such as power-law (PL), Band model, and cutoff power-law (CPL), and selected the one minimizing the Bayesian information criterion. The statistics and $\chi^2$ adopted to estimate the goodness of the spectral fits are severely affected by the low number of available spectral channels, but allow to provide a reliable modeling to calculate the corresponding flux in each time interval.

Spectra obtained for GRB~220101A in the different time intervals are shown in Fig.~\ref{SPECTRA}, together with the best-fit model and $1\sigma$ confidence region (lines and shaded regions). The main spectral parameters adopted for the best-fits are reported in Tab.~\ref{tabspectra}, together with fit statistics, and related fluences, evaluated in the 18~keV--10~MeV energy range and time-integrated on the corresponding time intervals. Hereafter, we briefly recap the most important points of the light-curve behavior and of the spectral evolution, throughout the different time intervals:

\begin{itemize}
    \item \textit{interval-A}: [$T_0$,$T_0+33$~s]. Duration: 33~s. This interval features the onset of the main episode of the prompt phase. In the SA X-ray energy range, the light-curve exhibits a slow rise, terminating with a sequence of emission spikes, and a peak emission flux reached at $T_0+29$~s, equal to $f_{p,A}^{SA}=(2.3\pm0.2)\cdot10^{-7}$~erg~cm$^{-2}$~s$^{-1}$. On the other hand, in the MCAL lowest-energy channels (175--1000~keV), it shows an initial ``hump'' between $T_0$ and $T_0+16$~s. The overall spectrum can be described by a Band model, with low-energy photon index $\alpha\sim-1$, high-energy photon index $\beta\sim-2.4$, and peak energy $E_{p,A}\sim280$~keV. A trial fit of this spectrum by means of a PL or a CPL model ends up with a $\chi^{2}_{red}\ge2$ and is therefore rejected. The related flux in this interval, estimated in the 18~keV--10~MeV energy range, is equal to $f_A=(1.19\pm0.12)\cdot10^{-6}$~erg~cm$^{-2}$~s$^{-1}$, resulting in a corresponding fluence $F_A=(3.93\pm0.39)\cdot10^{-5}$~erg~cm$^{-2}$ (both $90\%$ confidence level). In this interval, there is no evidence of significant emission above $\sim20$~MeV.
    
    \item \textit{interval-B}: [$T_0+33$~s,$T_0+58$~s]. Duration: 25~s. For the whole duration of the interval, the light-curve shows a sequence of spikes in the X-ray range. For what concerns the MCAL band, the event shows a sharp first peak, clearly visible from CH1 to CH6 (from 175~keV to 11.2~MeV), followed by a second longer and more spread emission episode. This interval represents the core emission of the burst and it can be clearly observed up to MCAL CH8 (i.e., 22.4--44.8~MeV). The related integrated spectrum can be modeled with a Band function with low-energy spectral index $\alpha\sim-0.3$, a rather flat high-energy spectral index $\beta\sim-2.2$, and peak energy $E_{p,B}\sim170$~keV. A fit of this spectrum with either a single PL or a CPL model produces a $\chi^{2}_{red}>2$ and is considered not reliable. In this interval, the flux in the 18~keV--10~MeV energy range is $f_B=(3.07\pm0.31)\cdot10^{-6}$~erg~cm$^{-2}$~s$^{-1}$, resulting in a corresponding fluence $F_B\sim(7.68\pm0.77)\cdot10^{-5}$~erg~cm$^{-2}$ (both $90\%$ confidence level). The overall flux in this interval is therefore more than twice that of \textit{interval-A}, with the peak flux of the whole burst encountered from $T_0+38$~s to $T_0+46$~s, equal to $f_{p,B}=(2.00\pm0.20)\cdot10^{-6}$~erg~cm$^{-2}$~s$^{-1}$. The corresponding time-integrated fluence results $F_{p,B}=(1.60\pm0.16)\cdot10^{-5}$~erg~cm$^{-2}$ and constitutes the release of about $12\%$ of the overall GRB fluence. In this interval, the right-hand side of the spectrum looks flatter, and exhibits a non-negligible emission up to $\sim50$~MeV. Considering a possible high-energy emission ($>50$~MeV) occurring at about $T_0+49$~s, potentially revealed in GRID RM data, and considering the typical spectral behavior often encountered in other remarkable bursts (e.g., GRB~090926A, GRB~130427A, GRB190114C), we investigated the possible existence of an additive spectral component, arising within this interval, and extending the spectrum up to the highest energies. From our data in the 18~keV--50~MeV energy range, there is no clear evidence of such an extra high-energy component. Nevertheless, \textit{interval-B} might be fitted by means of a Band model plus an additive PL with photon index $\sim1.35$, although resulting in a less reliable $\chi^2_{red}$ with respect to the one obtained for a simple Band model. We report the main spectral parameters of such fit in Tab.~\ref{tabspectra}.
    
    \item \textit{interval-C}: [$T_0+58$~s,$T_0+109$~s]. Duration: 51~s. This interval features the end of the prompt phase, with the light-curve decaying in intensity and exhibiting some last spiking both in the SA X-ray and in the MCAL high-energy range. Due to the low intensity, the spectrum is affected by large errors on the reconstructed flux. Nevertheless, we could fit this interval with a CPL, with photon index $\sim1.8$ and a cutoff energy $E_{c,C}\sim690$~keV, limiting the bulk of the emission spectrum to a synchrotron component in the sub-MeV domain. Another fit can be carried out by means of a single PL model with photon index $2.8$, although resulting in a worse $\chi^{2}_{red}=0.15$, with respect to the CPL model. The related flux in this interval, within a 18~keV--10~MeV energy range, is $f_C=(1.50\pm0.15)\cdot10^{-7}$~erg~cm$^{-2}$~s$^{-1}$, resulting in a corresponding fluence $F_C\sim(7.67\pm0.77)\cdot10^{-6}$~erg~cm$^{-2}$. The overall flux in \textit{interval-C} decays of more than one order of magnitude with respect to the former interval, setting off the end of the prompt phase. This interval is particularly relevant, as it features the end of the keV-MeV prompt emission, where high-energy GRBs (e.g., GRB~130427A, GRB~190114C) typically exhibit no noticeable spectral evolution above MeV - tens of MeV, and are well described by a single PL with photon index $\sim2$ \citep{Piron2016}. We notice that there could be some hints of $\sim$MeV emission above the corresponding cutoff energy $E_{c,C}$, but the very low-flux and limited count statistics prevent a reliable fit with an extra PL component, resulting in $\chi^2_{red}>2$. If a high-energy component is present and emerges in the final stages of the prompt phase, it shall mostly involve energies $\gtrsim50$~MeV or exhibit rather faint fluxes on the order of $\lesssim4-5\cdot10^{-7}$~erg~cm$^{-2}$~s$^{-1}$.
    
\end{itemize}

In the bottom block of Tab.~\ref{tabspectra}, we report also the best-fits obtained for intervals \textit{a1} and \textit{b1}, from the analysis of MCAL data. These intervals occur in between \textit{interval-A} and \textit{interval-B}, when the event doubles its flux and the overall spectrum becomes harder, enhancing the emission in the MeV - tens of MeV regime. \textit{Interval-A} and \textit{interval-B} can be described by Band models with peak energies ranging between 150--270~keV. As a consequence, the adopted Band functions would appear as simple PLs in the MCAL detector energy range, whose lower limit is set at 400~keV. For what concerns \textit{interval-a1}, we obtain a good best-fit using a PL with photon index $\sim2.5$. On the other hand, \textit{interval-b1} exhibits a flatter behavior and requires a higher photon index $\sim2.2$, compatible with the Band $\beta$ index obtained for \textit{interval-B} in the AGILE RMs.

\begin{figure*}
\centering
\includegraphics[width=0.8\linewidth]{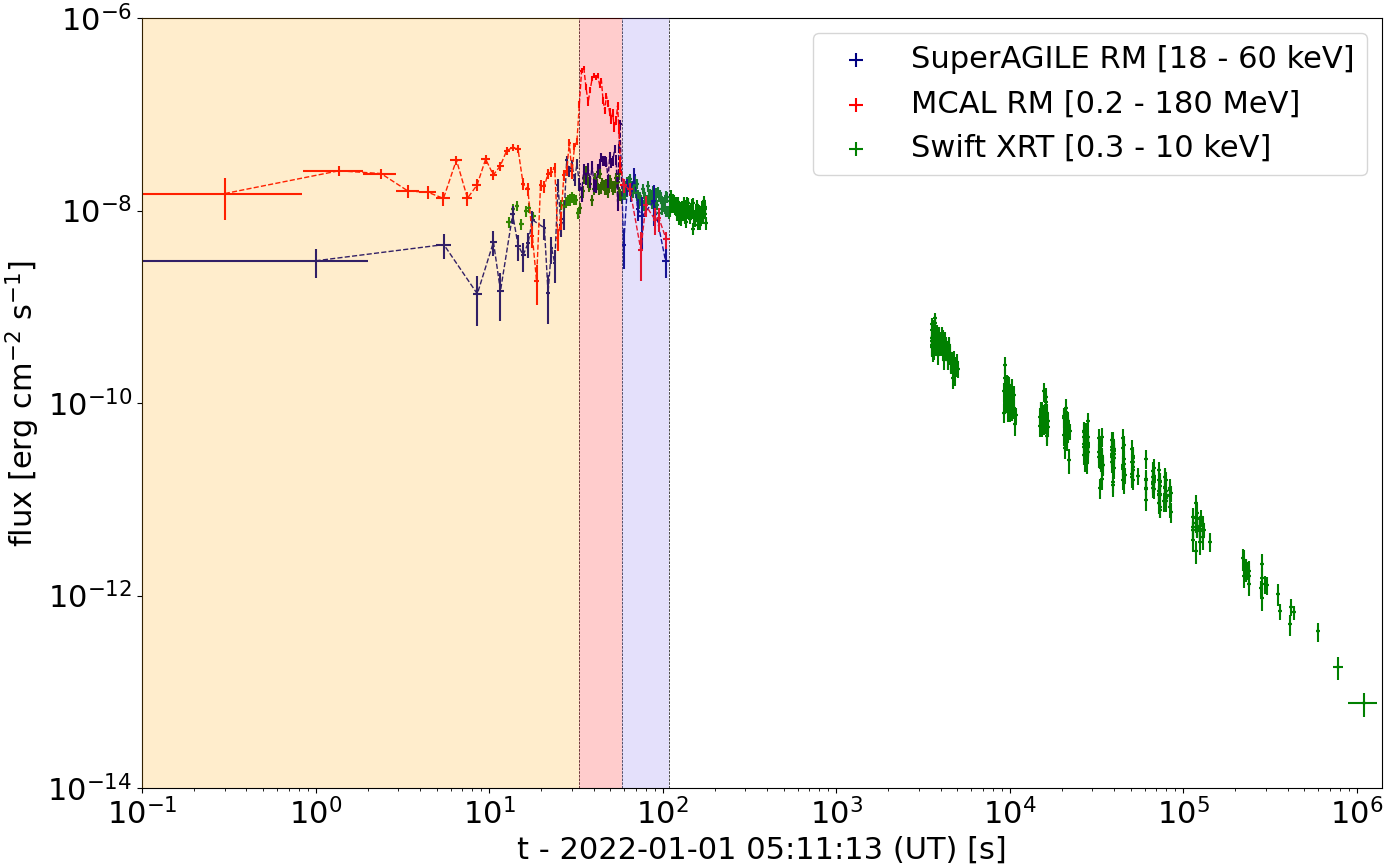}
\caption{GRB~220101A energy flux in the SuperAGILE (blue) and MCAL (red) energy ranges, evaluated by considering the best-fit spectra for intervals \textit{A} (yellow region), \textit{B} (red region), and \textit{C} (violet region), respectively. Swift XRT flux data (green) are also reported.}
\label{lcinflux}
\end{figure*}

The overall fluxes encountered in the different time intervals are relatively moderate, if compared with other remarkable bursts (e.g., GRB~130427A). Between $T_0$ and $T_0+109$~s, GRB~220101A exhibited a total average energy flux on the order of $\sim10^{-6}$~erg~cm$^{-2}$~s$^{-1}$. A fluence of $\sim10^{-4}$~erg~cm$^{-2}$ is obtained when integrating such flux over a relatively long time interval, mostly ascribed to time dilation effects due to the non-negligible redshift value. Fig.~\ref{lcinflux} shows the energy flux during the prompt emission, as detected by SA (blue) and MCAL (red), within 18--60~keV and 0.2-180~MeV, respectively. Moreover, we report the publicly available Swift XRT data\footnote{https://www.swift.ac.uk/xrt\_curves/0} (green), in the 0.3-10~keV energy range, which cover both the prompt and the extended emission phase for times greater than 109~s. It is interesting to notice that for intervals \textit{A} and \textit{B}, the flux above $\sim200$~keV is dominant, whereas in the latter \textit{interval-C}, it fades revealing the X-ray flux in the keV range, which persists up to very late times, setting the onset of the afterglow phase.

\subsection{Energetics}

\begin{figure}
\centering
\includegraphics[width=1.0\linewidth]{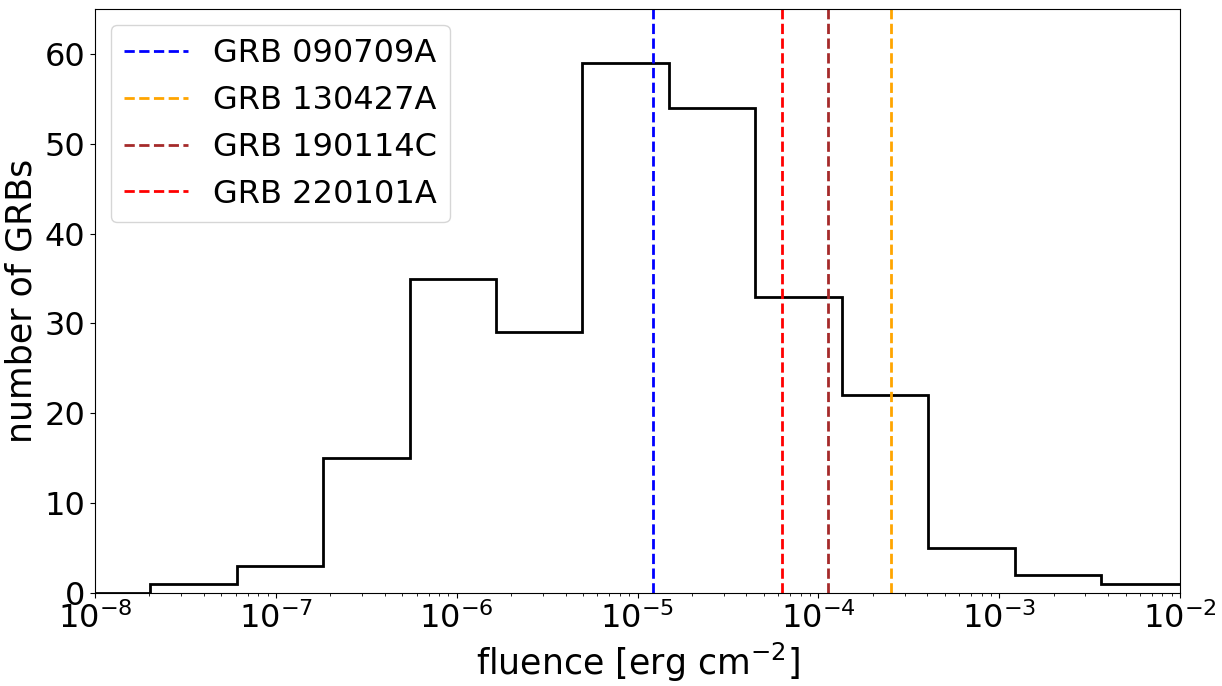}
\caption{Fluence distribution of the 258~bursts of the second MCAL GRB catalog. Fluences are evaluated in the 400~keV--10~MeV energy range. It can be noticed that GRB~220101A (red dashed line) is placed within the first quartile of GRB fluences among the bursts detected by MCAL. For comparison, fluences of other remarkable bursts are reported with colored dashed lines.}
\label{fluence}
\end{figure}

\begin{figure*}
\centering
\includegraphics[width=1.0\linewidth]{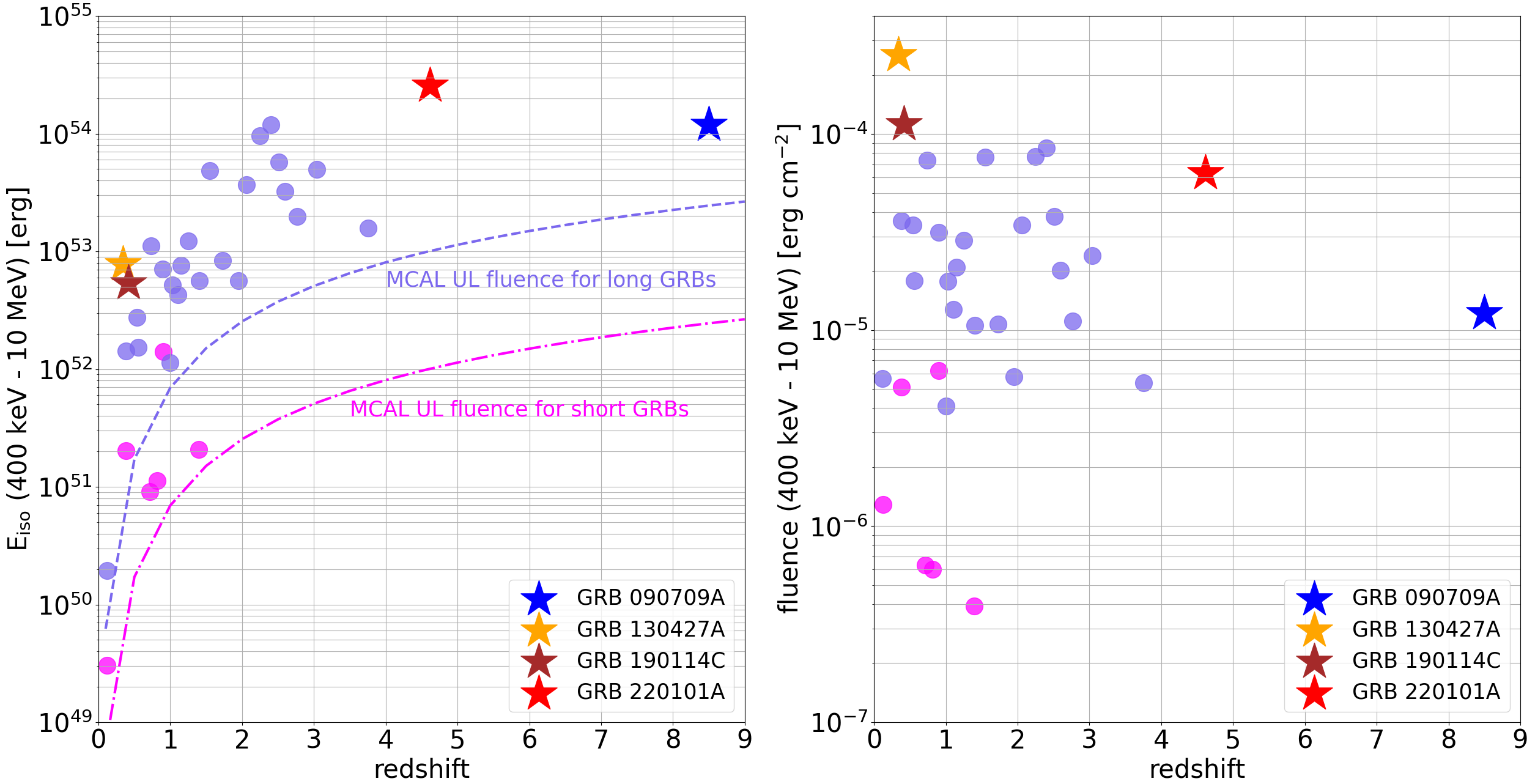}
\caption{Left: Reconstructed isotropic energy, in the 400~keV--10~MeV energy range, plotted with respect to redshift, for 32 GRBs detected by the AGILE MCAL. The dot-dashed magenta and dashed blue lines represent average MCAL upper limit fluences, necessary to trigger on short (magenta dots) and long (blue dots) GRBs, respectively. Stars represent some remarkable GRBs, which exhibited either high fluences or high redshift values. It can be noticed that GRB~220101A (red star) has the second higher redshift in the MCAL GRB sample, as well as the highest $E_{iso}$ among the MCAL detected bursts, for which a redshift was available. Right: MCAL GRB fluence, evaluated in the 400~keV--10~MeV energy range, plotted with respect to redshift, for the same events. GRB~220101A fluence is placed within the first quartile among MCAL GRBs.}
\label{redshift}
\end{figure*}

The overall prompt emission of GRB~220101A, evaluated from $T_0$ to $T_0+109$~s, can be fitted with a Band model, resulting in a flux $f=(1.55\pm0.16)\cdot10^{-6}$~erg~cm$^{-2}$~s$^{-1}$. The corresponding fluence, integrated on the total 109~s time duration, is equal to $F=(1.69\pm0.17)\cdot10^{-4}$~erg~cm$^{-2}$: such value is a rather typical value among the brightest GRBs detected by MCAL \citep{Ursi2022catalog}, and places GRB~220101A within the first quartile of GRB fluences in the MCAL burst catalog, as shown in Fig~\ref{fluence}. Considering the 400~keV--10~MeV energy range, which is more suitable for MCAL analysis, the burst exhibits a flux $f[0.4-10$~MeV$]=(6.34\pm0.64)\cdot10^{-7}$~erg~cm$^{-2}$~s$^{-1}$ and a fluence $F[0.4-10$~MeV$]=(6.91\pm0.69)\cdot10^{-5}$~erg~cm$^{-2}$. Assuming the redshift z=4.62 \citep{Fu2022,Perley2022,Fynbo2022,Tomasella2022} and a standard cosmological model provided by \cite{Ade2014} (i.e., $H_0=67.3$~km~s$^{-1}$~Mpc$^{-1}$, $\Omega_M=0.315$, and $\Omega_{\Lambda}=0.685$), we end up with an isotropic energy release and a peak luminosity in the rest-frame equal to $E_{iso}=2.54\cdot10^{54}$~erg and $L_{iso}=2.34\cdot10^{52}$~erg~s$^{-1}$, respectively. This represents the highest value of $E_{iso}$ encountered among the GRBs collected by AGILE to date. The left panel of Fig.~\ref{redshift} shows the reconstructed equivalent isotropic energy with respect to redshift, for 32~GRBs detected by the AGILE MCAL for which a redshift was provided by X-ray or optical observations of their afterglows. The $E_{iso}$ is evaluated in the 400~keV--10~MeV energy range. For 8 GRBs, we fitted the corresponding GRB spectrum with a Band model with peak energy $>400$~keV, whereas in the remaining 24 cases, we fitted the spectrum with a simple PL, corresponding to the right-hand side of the corresponding Band, or CPL model, describing the burst spectrum. In all these cases, the 400~keV--10~MeV energy range allows us to investigate these spectra in the MCAL data, and to provide a consistent description of the energetics of the burst sample. In figure, the dashed and dash-dotted lines represent the average MCAL upper limit (UL) fluences, corresponding to the onboard trigger thresholds for the detection of long (blue dots) and short (magenta dots) GRBs, respectively. It can be noticed that, even in the relatively high energy range considered for this analysis, GRB~220101A (red star) is the event with the highest $E_{iso}$ among the MCAL GRBs, even higher than other remarkable bursts, such as GRB~090709A (blue star), GRB~130427A (orange star), and GRB~190114C (magenta star). The right panel of Fig.~\ref{redshift} shows the GRB fluence, evaluated in the 400~keV--10~MeV energy range, for the same 32 bursts detected by MCAL. We notice that GRB~220101A is the second farthest event detected by AGILE, whereas it does not exhibit a particularly large fluence, which has a quite standard value among the GRBs detected by MCAL. As a consequence, the major cause of its very large reconstructed isotropic energy release is ascribed to the large distance at which the event occurred.

\section{The afterglow of GRB~220101A}

We carried out a theoretical analysis of the afterglow emission, based on the forward shock model, consisting in the production of photons via synchrotron and Inverse Compton (IC) mechanisms, due to the interaction of the expanding shock with particles of the surrounding medium. The shock expansion, as shown in \cite{Blandford1976}, can be adiabatic, if the shock expands mantaining a constant internal energy, or radiative, if there is a high efficiency in converting the shock internal energy into radiated energy. The latter evolution type can be verified in the prompt, or in the early stages of the GRB afterglow emission, which correspond to phases during which the outflow produced by the GRB progenitor has a large amount of energy and the system can radiate very efficiently. At later stages of the afterglow, the photon production rate decreases. In the following treatment, we analyze the afterglow emission considering the adiabatic evolution, cross-checking our prediction using Swift XRT data. 

\begin{figure*}[]
\centering
\includegraphics[width=1.0\linewidth]{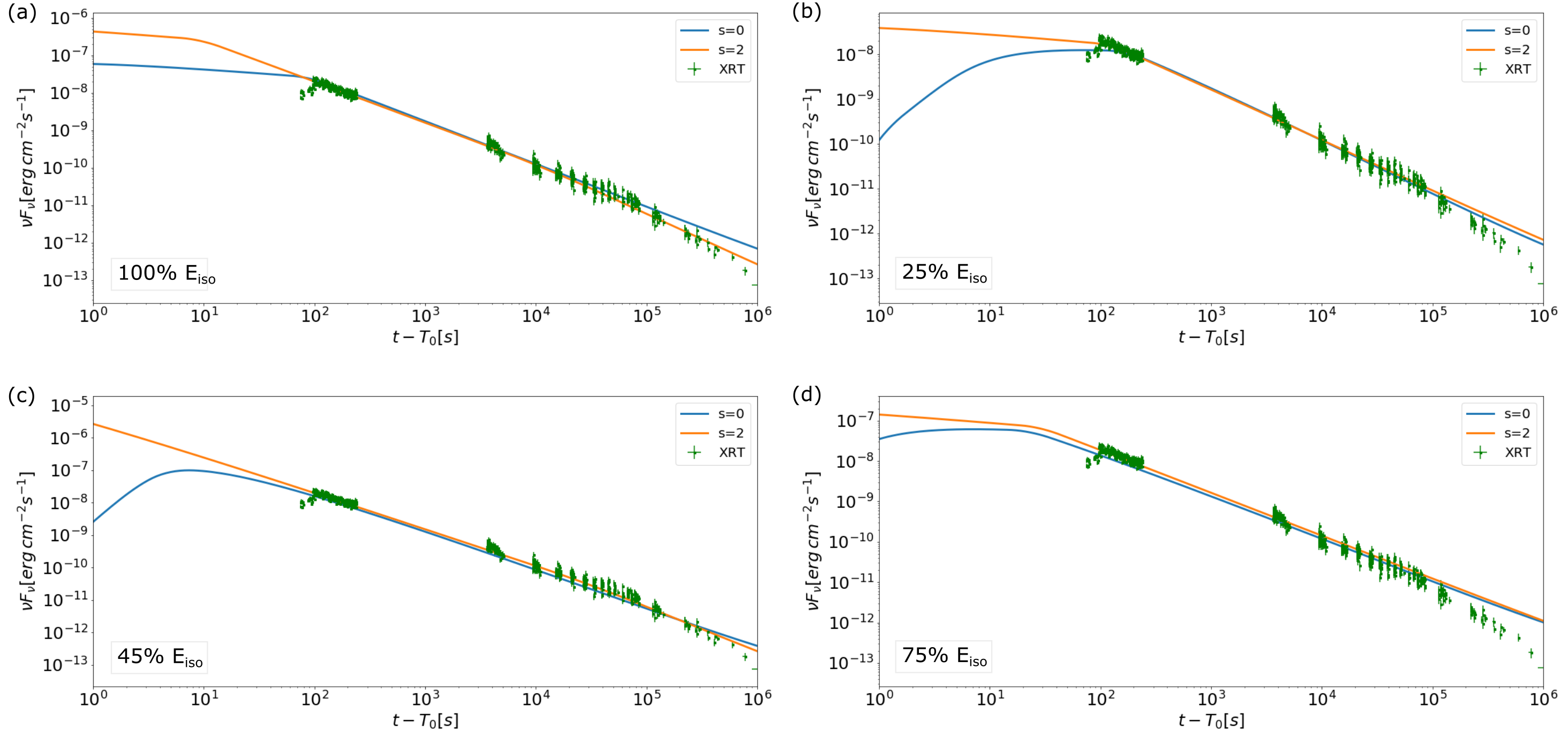}
\caption{Expected afterglow light-curves for the homogeneous (blue line) and wind-like (orange line) scenarios, considering (a) the entire $E_{iso}$, (b) $25\%$ of the $E_{iso}$, (c) $45\%$ of the $E_{iso}$, and (d) $75\%$ of the $E_{iso}$. Green squares represent the Swift XRT data. Here, $T_0$ is the Swift $T_0$= 2022-01-01 05:10:11 UT.}
\label{GRB220101A_XRT_plot_light_curve}
\end{figure*}

Tab.~\ref{tab:parameters} reports the parameters adopted for our analysis. It can be noticed that the very high $E_{iso}=3.64\cdot10^{54}$~erg
\citep[as reported by][]{Atteia2022,Tsvetkova2022a,Ruffini2022a} strongly limits the initial value of the critical frequencies $\nu_m$ and $\nu_c$, which depend on the evolution type and on the external density profile, as shown in Tab.~\ref{tab:critical_frequencies}. The $\nu_m$ and $\nu_c$ are the synchrotron frequencies evaluated for two peculiar electron Lorentz factors, denoted as $\gamma_m$ and $\gamma_c$: the former corresponds to the minimum value from which the electron distribution is defined, whereas the latter is the Lorentz factor of an electron which could radiate with energy equal to $m_ec^2$. These electron Lorentz factors are not fixed for each GRB: $\gamma_m$ and $\gamma_c$, and consequently the associated synchrotron frequencies $\nu_m$ and $\nu_c$, show a temporal behavior that depends on the shock evolution and by the environment in which the shock propagates. The main consequence is that the afore-mentioned critical frequencies could cross different detector bands at different times, due to their temporal dependency. In the homogeneous scenario, the XRT band ($0.3-10$~keV) is initially placed between those frequencies. The predicted light-curve (blue line in Fig.\ref{GRB220101A_XRT_plot_light_curve} (a)) has only one temporal break, whose nature is related to the crossing of the $\nu_m$ in the XRT band. After this break, the light-curve behaves as a power-law, persisting for the whole afterglow emission. In Fig.~\ref{GRB220101A_XRT_plot_light_curve}, the reported $T_0$ is the Swift $T_0$= 2022-01-01 05:10:11 (UT). This behavior is compatible with XRT data until $\sim 4\cdot10^{4}$~s where a new temporal break occurs. The prediction model improves if we adopt a wind-like scenario. The expected light-curve (orange line in Fig.~\ref{GRB220101A_XRT_plot_light_curve} (a)) shows a break either at early times, when the break frequency $\nu_m$ crosses the observed frequency $\nu$, and at later times, when the break frequency $\nu_c$ crosses the observed frequency $\nu$. This behavior is similar to that of GRB~190114C, reported by \cite{MAGIC2019}, \cite{Ajello2020}, and \cite{Ursi2020magic}. It is important to notice that the forward shock model is useful for the study of the pure afterglow emission; the prompt phase of GRB~220101A lasted for a very long time interval ($\sim 100$~s) and this implies that the model used for this analysis becomes physically relevant after 100~s since the Swift $T_0$. Although in Fig.~\ref{GRB220101A_XRT_plot_light_curve} (a) there are temporal breaks for $t\leq 100$~s, when the afterglow is not present yet, a discussion on the nature of these breaks is useful for the comprehension of how the forward shock model predicts the various temporal breaks.

The lack of a late temporal break for the homogeneous scenario is linked to the temporal evolution of the $\nu_c$, as well as to the dependency on the $E_{iso}$. We can notice from Tab.~\ref{tab:critical_frequencies} that $\nu_c$ decreases with time and, for higher values of $E_{iso}$, the $\nu_c$ exhibits a lower initial value: this latter effect makes the $\nu_c$ lay from the beginning in the lower part of the spectrum with respect to the XRT band. As a consequence, if at early times the $\nu_c$ lays already below the XRT band, this critical frequency could not cross the detector band. On the other hand, in the wind-like scenario, this transition may occur, as $\nu_c$ increases with time with a velocity that is lower than the decreasing of $\nu_m$. This could explain why the crossing of $\nu_m$ generally takes place at early times, while the crossing of $\nu_c$ takes place at late times.

\begin{table}[]
\centering
\caption{Parameters for homogeneous and wind-like scenarios}
\begin{center}
\begin{tabular}{|c|c|c|c|c|c|c|c|c|}
\hline 
$E_{iso}$ & scen. & $E_{53}$ & $p$ & $\Gamma_{0}$ & $\epsilon_{e;-1}$ & $\epsilon_{B;-3}$ & $n_{0}$ & $A_{*}$ \\ \hline \hline
\multirow{2}{*}{$100\%$} & hom. & 36.4 & 2.20 & 1500 & 4 & 60 & 0.08 & -- \\ \cline{2-9}
 & wind & 36.4 & 2.15 & 700 & 5.5 & 0.05 & -- & 16 \\ \hline
\multirow{2}{*}{$75\%$} & hom. & 27.3 & 2.10 & 200 & 3 & 60 & 0.5 & -- \\ \cline{2-9}
 & wind & 27.3 & 2.10 & 200 & 3.55 & 55 & -- & 9 \\ \hline 
\multirow{2}{*}{$45\%$} & hom. & 17.0 & 2.30 & 200 & 0.4 & 6 & 0.8 & -- \\ \cline{2-9}
 & wind & 15.0 & 2.20 & 200 & 0.55 & 0.05 & -- & 16 \\ \hline 
\multirow{2}{*}{$25\%$} & hom. & 11.0 & 2.30 & 150 & 7.0 & 60 & 0.8 & -- \\ \cline{2-9}
 & wind & 10.0 & 2.20 & 150 & 7.5 & 50 & -- & 16 \\ \hline 
\end{tabular}
\end{center}
Parameters of GRB~220101A for a progenitor evolving in a homogeneous, or in a wind-like scenario. For each case, we considered different fractions of the total $E_{iso}$, that are converted into kinetic energy of the shock.
\label{tab:parameters}
\end{table}

\begin{table}[]
\caption{Critical frequencies in the adiabatic evolution}
\centering
\begin{center}
\begin{tabular}{|c||c|c|}
\hline
scenario & $\nu_{m}$ & $\nu_{c}$ \\ \hline \hline
homogeneous & $E_{iso}^{1/2}$, $t^{-3/2}$ & $E_{iso}^{-1/2}$, $t^{-1/2}$ \\ \hline
wind-like & $E_{iso}^{1/2}$, $t^{-3/2}$ & $E_{iso}^{1/2}$, $t^{1/2}$ \\ \hline
\end{tabular}
\end{center}
Dependencies of the $\nu_m$ and $\nu_c$ critical frequencies on the isotropic energy and on the time after the trigger time, for a GRB progenitor evolving in a homogeneous or wind-like circumburst density profile
\label{tab:critical_frequencies}
\end{table}

\subsection{Parameter discussion}

The equipartition parameters $\epsilon_e$ and $\epsilon_B$ respectively represent the fraction energy of the electron population and of the magnetic field, with respect to the energy of the shock. As shown in Tab.~\ref{tab:parameters}, the best fit obtained for $100\%$ of $E_{iso}$ exhibits a very large $\Gamma_0$ initial shock Lorentz factor, with respect to the typical values characterizing other remarkable and intrinsically more energetic GRBs \citep[e.g., the GRB~190114C, whose afterglow was modeled with a $\Gamma_0\lesssim700$, or even with a $\Gamma_0\lesssim450$][]{MAGIC2019,Derishev&Piran2021}. A Lorentz factor of 1500 would therefore make difficult to explain a high efficiency capable of converting kinetic energy into radiation, as well as the presence of an intense magnetic field. As a consequence, we made the hypothesis that not the entire energy released by the GRB progenitor is converted into the shock kinetic energy: we assume that only a fraction of $E_{iso}$ is relevant for the origin of a primary shock, which then interacts with the external medium, producing the observed emission. In order to evaluate this, we carried out some tests by adopting a $25\%$, $45\%$ and $75\%$ of the total $E_{iso}$, whose light-curves are reported in panels (b), (c), and (d) of Fig.~\ref{GRB220101A_XRT_plot_light_curve}. The analyzed cases are for:

\begin{itemize}

\item $25\%$ of the $E_{iso}$, Fig.~\ref{GRB220101A_XRT_plot_light_curve} (b). In this case, the predicted light-curve for a homogeneous scenario (s=0) has two near temporal breaks at very early times, due to the crossing of $\nu_m$ and of $\nu_c$, respectively. The first break, as previously pointed out, is situated at $t<100$~s, when the afterglow emission has still not emerged. The second one occurs within the valid region, but it seems not compatible with the early XRT data. Successively, the light-curve evolves as a power-law with good agreement with the XRT trend, up to $\sim4\cdot10^{4}$~s, after which the XRT data diverge from the model slope. The discussion for the wind-like scenario (s=2) is similar: however, in this case, it shows only one break due to the transition of $\nu_m$ over the XRT band, consistent with the XRT data.

\item $45\%$ of the $E_{iso}$, Fig.~\ref{GRB220101A_XRT_plot_light_curve} (c). Also in this case, the homogenenous scenario foresees the existence of an early break regarding the early transition of $\nu_c$, which does not lay within the valid region. As a consequence, we only treated the wind-like scenario, which does not imply any early break (the transition of $\nu_m$ occurs few milliseconds after the trigger time, in the not-consistent region).

\item $75\%$ of the $E_{iso}$, Fig.~\ref{GRB220101A_XRT_plot_light_curve} (d). The obtained fit is similar to that for $25\%$ of $E_{iso}$, although exhibiting different light-curve shapes and different temporal breaks.

\end{itemize}

From this test, we notice that the cases adopting the 25\% and 75\% of the $E_{iso}$ are in agreement with the XRT data until $4\cdot10^{4}$~s; however, after this break, the temporal slopes of the expected light curves do not follow the trend of the observed light curves. On the other hand, the cases adopting $100\%$ and $45\%$ of the $E_{iso}$ show a good compatibility with the observed data, either before and after the break time. Moreover, both these best fits exhibit a $\Gamma_0>100$, which is typically required to have an optically thin medium, allowing the gamma-ray emissions to escape the system \citep{Piran1999}. However, as already pointed out in this section, the very large value of $\Gamma_0$ resulting from the case with $100\%$ of $E_{iso}$ makes us rule out this best fit as a possible modeling for the burst afterglow. We therefore consider the case with $45\%$ of $E_{iso}$ as the best configuration to describe the shock. From a physical point of view, this means that the progenitor of GRB~220101A released a large amount of energy, and that only a considerable fraction of it is actually converted into kinetic energy, necessary for the forward shock expansion and for the beginning of the afterglow phase. This does not exclude that multiple shocks may be generated, although the primary one carries out the major part of the event energy, producing most of the detected emission. Moreover, in both configurations, the wind-like density profile is the scenario providing the better fits for the Swift XRT data of the afterglow emission; the slope before and after the late break is compatible with the observed ones and the position of the temporal break is consistent with the observed data. We can conclude that the GRB~220101A evolved with high probability in a wind-like density medium, and that the energy carried away by the shock is almost half of the entire $E_{iso}$ released by the event.

\begin{figure*}
\centering
  \includegraphics[width=0.8\linewidth]{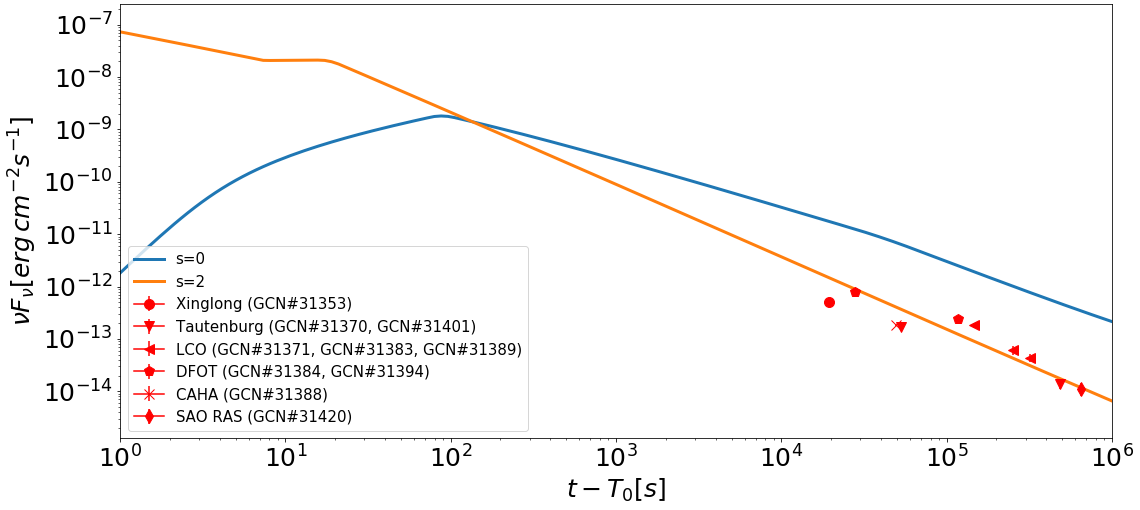}
\caption{Predicted light-curves in the homogeneous (blue line) and wind-like (orange line) scenario, considering a $45\%$ of the total burst $E_{iso}$, and the optical data (red points) retrieved in public GCN circulars (reported in the legend).}
\label{GRB220101A_Optical_plot_light_curve}
\end{figure*}

We cross-checked the results obtained from our analysis with the optical emission in the R-band, considering the preliminary fluxes reported in various GCN by optical observatories, such as Xinglong \citep{Fu2022}, Tautenburg \citep{NicuesaGuelbenzu2022a,NicuesaGuelbenzu2022b}, LCO \citep{Strausbaugh2022a,Strausbaugh2022b,Strausbaugh2022c}, DFOT \citep{Dimple2022,Ror2022}, CAHA \citep{Caballero-Garcia2022}, and SAO RAS \citep{Moskvitin2022}. As shown in Fig.~\ref{GRB220101A_Optical_plot_light_curve}, we notice that the optical fluxes show a good agreement with the model previously obtained by considering a $45\%$ of the total $E_{iso}$, and an expansion shock evolving in a wind-like scenario (s=2).

\section{Conclusions}

GRB~220101A represents a record burst, exhibiting one of the largest isotropic energy releases ever observed from a GRB \citep[$E_{iso}\sim3.64\cdot10^{54}$~erg, as reported by][]{Atteia2022,Tsvetkova2022a}. Taking advantage of the SuperAGILE, Anti-Coincidence, and MCAL data, acquired by the onboard scientific ratemeters, the AGILE satellite provides a broad-band overall description of the evolution of this GRB, from few tens of keV to tens of MeV, highlighting some peculiar features of its temporal and spectral behavior. The event lasted more than 100~s and its localization region was fully inside the AGILE field of view for most of its duration. We divided the prompt emission into three main intervals: the onset of the burst dominated by X-ray/soft gamma-ray emission, a central interval where the burst reaches its peak flux (especially in the MeV energy range), and a final interval with a fading low-flux emission, setting off the end of the prompt phase. The average fluxes encountered in the different time intervals are relatively ordinary, with respect to those observed in other remarkable GRBs, exhibiting values around $\sim10^{-6}$~erg~cm$^{-2}$~s$^{-1}$. The central interval between $T_0+38$~s and $T_0+46$~s features the burst peak emission, where about $12\%$ of the overall fluence is released. The central and final intervals of the GRB prompt might feature an additive power-law component, extending the spectrum to several tens of MeV, although our data do not allow to carry out a detailed analysis, due to the low-statistics. Nevertheless, a good fit in the 18~keV--50~MeV does not require extra components beyond single Band or cutoff power-law models.

The analysis of the afterglow of GRB~220101A, carried out using the public Swift XRT data and adopting the forward shock model, reveals that the surrounding environment in which the event took place is mainly compatible with a wind-like density profile. The progenitor of GRB 220101A is therefore likely a massive star, that provided a wind-like environment for its circumburst medium density. A best-fit for this modeling is obtained considering that only a fraction ($\sim45\%$) of the entire $E_{iso}$ is converted into kinetic energy, necessary for the expansion of the forward shock. Such model shows a very good agreement also with the optical data retrieved from the preliminary analysis of ground observatories, reported in public GCN notices.

In the 400~keV--10~MeV energy range, GRB~220101A does not exhibit a particularly remarkable fluence, whose value ranks within the first quartile among the fluences of the GRBs detected by MCAL. However, this event has the second highest redshift among the MCAL bursts, and due to the huge distance of the progenitor, it exhibits a reconstructed equivalent isotropic energy equal to $E_{iso}=2.54\cdot10^{54}$~erg (0.4-10~MeV), becoming the most energetic event of the whole MCAL GRB sample collected to date.

\acknowledgements

AGILE is a mission of the Italian Space Agency (ASI), with coparticipation of INAF (Istituto Nazionale di Astrofisica) and INFN (Istituto Nazionale di Fisica Nucleare). This work was carried out in the frame of the ASI-INAF agreement I/028/12/6. This work makes use of Swift public data, available at the https://www.swift.ac.uk/repository. This work reports data published by Xinglong, Tautenburg, LCO, DFOT, CAHA, SAO RAS observatories.

\newpage


\begin{thebibliography}{}
\expandafter\ifx\csname natexlab\endcsname\relax\def\natexlab#1{#1}\fi
\providecommand{\url}[1]{\href{#1}{#1}}
\providecommand{\dodoi}[1]{doi:~\href{http://doi.org/#1}{\nolinkurl{#1}}}
\providecommand{\doeprint}[1]{\href{http://ascl.net/#1}{\nolinkurl{http://ascl.net/#1}}}
\providecommand{\doarXiv}[1]{\href{https://arxiv.org/abs/#1}{\nolinkurl{https://arxiv.org/abs/#1}}}

\bibitem[{{Ackermann} {et~al.}(2010){Ackermann}, {Asano}, {Atwood}, {Axelsson},
  {Baldini}, {Ballet}, {Barbiellini}, {Baring}, {Bastieri}, {Bechtol},
  {Bellazzini}, {Berenji}, {Bhat}, {Bissaldi}, {Blandford}, {Bloom},
  {Bonamente}, {Borgland}, {Bouvier}, {Bregeon}, {Brez}, {Briggs}, {Brigida},
  {Bruel}, {Buson}, {Caliandro}, {Cameron}, {Caraveo}, {Carrigan},
  {Casandjian}, {Cecchi}, {{\c{C}}elik}, {Charles}, {Chiang}, {Ciprini},
  {Claus}, {Cohen-Tanugi}, {Connaughton}, {Conrad}, {Dermer}, {de Palma},
  {Dingus}, {Silva}, {Drell}, {Dubois}, {Dumora}, {Farnier}, {Favuzzi},
  {Fegan}, {Finke}, {Focke}, {Frailis}, {Fukazawa}, {Fusco}, {Gargano},
  {Gasparrini}, {Gehrels}, {Germani}, {Giglietto}, {Giordano}, {Glanzman},
  {Godfrey}, {Granot}, {Grenier}, {Grondin}, {Grove}, {Guiriec}, {Hadasch},
  {Harding}, {Hays}, {Horan}, {Hughes}, {J{\'o}hannesson}, {Johnson}, {Kamae},
  {Katagiri}, {Kataoka}, {Kawai}, {Kippen}, {Kn{\"o}dlseder}, {Kocevski},
  {Kouveliotou}, {Kuss}, {Lande}, {Latronico}, {Lemoine-Goumard}, {Llena
  Garde}, {Longo}, {Loparco}, {Lott}, {Lovellette}, {Lubrano}, {Makeev},
  {Mazziotta}, {McEnery}, {McGlynn}, {Meegan}, {M{\'e}sz{\'a}ros}, {Michelson},
  {Mitthumsiri}, {Mizuno}, {Moiseev}, {Monte}, {Monzani}, {Moretti},
  {Morselli}, {Moskalenko}, {Murgia}, {Nakajima}, {Nakamori}, {Nolan},
  {Norris}, {Nuss}, {Ohno}, {Ohsugi}, {Omodei}, {Orlando}, {Ormes}, {Ozaki},
  {Paciesas}, {Paneque}, {Panetta}, {Parent}, {Pelassa}, {Pepe},
  {Pesce-Rollins}, {Piron}, {Preece}, {Rain{\`o}}, {Rando}, {Razzano},
  {Razzaque}, {Reimer}, {Ritz}, {Rodriguez}, {Roth}, {Ryde}, {Sadrozinski},
  {Sander}, {Scargle}, {Schalk}, {Sgr{\`o}}, {Siskind}, {Smith}, {Spandre},
  {Spinelli}, {Stamatikos}, {Stecker}, {Strickman}, {Suson}, {Tajima},
  {Takahashi}, {Takahashi}, {Tanaka}, {Thayer}, {Thayer}, {Thompson},
  {Tibaldo}, {Toma}, {Torres}, {Tosti}, {Tramacere}, {Uchiyama}, {Uehara},
  {Usher}, {van der Horst}, {Vasileiou}, {Vilchez}, {Vitale}, {von Kienlin},
  {Waite}, {Wang}, {Wilson-Hodge}, {Winer}, {Wu}, {Yamazaki}, {Yang}, {Ylinen},
  \& {Ziegler}}]{Ackermann2010}
{Ackermann}, M., {Asano}, K., {Atwood}, W.~B., {et~al.} 2010, \apj, 716, 1178,
  \dodoi{10.1088/0004-637X/716/2/1178}

\bibitem[{{Ackermann} {et~al.}(2013){Ackermann}, {Ajello}, {Asano}, {Axelsson},
  {Baldini}, {Ballet}, {Barbiellini}, {Bastieri}, {Bechtol}, {Bellazzini},
  {Bhat}, {Bissaldi}, {Bloom}, {Bonamente}, {Bonnell}, {Bouvier}, {Brand t},
  {Bregeon}, {Brigida}, {Bruel}, {Buehler}, {Burgess}, {Buson}, {Byrne},
  {Caliandro}, {Cameron}, {Caraveo}, {Cecchi}, {Charles}, {Chaves},
  {Chekhtman}, {Chiang}, {Chiaro}, {Ciprini}, {Claus}, {Cohen-Tanugi},
  {Connaughton}, {Conrad}, {Cutini}, {D'Ammand o}, {de Angelis}, {de Palma},
  {Dermer}, {Desiante}, {Digel}, {Dingus}, {Di Venere}, {Drell},
  {Drlica-Wagner}, {Dubois}, {Favuzzi}, {Ferrara}, {Fitzpatrick}, {Foley},
  {Franckowiak}, {Fukazawa}, {Fusco}, {Gargano}, {Gasparrini}, {Gehrels},
  {Germani}, {Giglietto}, {Giommi}, {Giordano}, {Giroletti}, {Glanzman},
  {Godfrey}, {Goldstein}, {Granot}, {Grenier}, {Grove}, {Gruber}, {Guiriec},
  {Hadasch}, {Hanabata}, {Hayashida}, {Horan}, {Hou}, {Hughes}, {Inoue},
  {Jackson}, {Jogler}, {J{\'o}hannesson}, {Johnson}, {Johnson}, {Kamae},
  {Kataoka}, {Kawano}, {Kippen}, {Kn{\"o}dlseder}, {Kocevski}, {Kouveliotou},
  {Kuss}, {Lande}, {Larsson}, {Latronico}, {Lee}, {Longo}, {Loparco},
  {Lovellette}, {Lubrano}, {Massaro}, {Mayer}, {Mazziotta}, {McBreen},
  {McEnery}, {McGlynn}, {Michelson}, {Mizuno}, {Moiseev}, {Monte}, {Monzani},
  {Moretti}, {Morselli}, {Murgia}, {Nemmen}, {Nuss}, {Nymark}, {Ohno},
  {Ohsugi}, {Omodei}, {Orienti}, {Orlando}, {Paciesas}, {Paneque}, {Panetta},
  {Pelassa}, {Perkins}, {Pesce-Rollins}, {Piron}, {Pivato}, {Porter}, {Preece},
  {Racusin}, {Rain{\`o}}, {Rando}, {Rau}, {Razzano}, {Razzaque}, {Reimer},
  {Reimer}, {Reposeur}, {Ritz}, {Romoli}, {Roth}, {Ryde}, {Saz Parkinson},
  {Schalk}, {Sgr{\`o}}, {Siskind}, {Sonbas}, {Spandre}, {Spinelli}, {Suson},
  {Tajima}, {Takahashi}, {Takeuchi}, {Tanaka}, {Thayer}, {Thayer}, {Thompson},
  {Tibaldo}, {Tierney}, {Tinivella}, {Torres}, {Tosti}, {Troja}, {Tronconi},
  {Usher}, {Vandenbroucke}, {van der Horst}, {Vasileiou}, {Vianello}, {Vitale},
  {von Kienlin}, {Winer}, {Wood}, {Wood}, {Xiong}, \& {Yang}}]{Ackermann2013}
{Ackermann}, M., {Ajello}, M., {Asano}, K., {et~al.} 2013, \apjs, 209, 11,
  \dodoi{10.1088/0067-0049/209/1/11}

\bibitem[{Ajello {et~al.}(2020)Ajello, Arimoto, Axelsson, Baldini, Barbiellini,
  Bastieri, Bellazzini, Berretta, Bissaldi, Blandford, Bonino, Bottacini,
  Bregeon, Bruel, Buehler, Burns, Buson, Cameron, Caputo, Caraveo, Cavazzuti,
  Chen, Chiaro, Ciprini, Cohen-Tanugi, Costantin, Cutini, D'Ammando, DeKlotz,
  de~la Torre~Luque, de~Palma, Desai, Lalla, Venere, Dirirsa, Fegan,
  Franckowiak, Fukazawa, Funk, Fusco, Gargano, Gasparrini, Giglietto, Gill,
  Giordano, Giroletti, Granot, Green, Grenier, Grondin, Guiriec, Hays, Horan,
  J{\'{o}}hannesson, Kocevski, Kovac'evic', Kuss, Larsson, Latronico,
  Lemoine-Goumard, Li, Liodakis, Longo, Loparco, Lovellette, Lubrano, Maldera,
  Malyshev, Manfreda, Mart{\'{\i}}-Devesa, Mazziotta, McEnery, Mereu, Meyer,
  Michelson, Mitthumsiri, Mizuno, Monzani, Moretti, Morselli, Moskalenko,
  Negro, Nuss, Omodei, Orienti, Orlando, Palatiello, Paliya, Paneque, Pei,
  Persic, Pesce-Rollins, Petrosian, Piron, Poon, Porter, Principe, Racusin,
  Rain{\`{o}}, Rando, Rani, Razzano, Razzaque, Reimer, Reimer, Ryde, Parkinson,
  Serini, Sgr{\`{o}}, Siskind, Spandre, Spinelli, Tajima, Takagi, Takahashi,
  Tak, Thayer, Thompson, Torres, Troja, Valverde, Klaveren, Wood, Yassine,
  Zaharijas, Mailyan, Bhat, Briggs, Cleveland, Giles, Goldstein, Hui,
  Malacaria, Preece, Roberts, Veres, Wilson-Hodge, von Kienlin, Cenko, O'Brien,
  Beardmore, Lien, Osborne, Tohuvavohu, D'Elia, D'A{\`{\i}}, Perri, Gropp,
  Klingler, Capalbi, Tagliaferri, Stamatikos, \& Pasquale}]{Ajello2020}
Ajello, M., Arimoto, M., Axelsson, M., {et~al.} 2020, The Astrophysical
  Journal, 890, 9, \dodoi{10.3847/1538-4357/ab5b05}

\bibitem[{{Arimoto} {et~al.}(2022){Arimoto}, {Scotton}, {Longo}, \& {Fermi-LAT
  Collaboration}}]{Arimoto2022}
{Arimoto}, M., {Scotton}, L., {Longo}, F., \& {Fermi-LAT Collaboration}. 2022,
  GRB Coordinates Network, 31350, 1

\bibitem[{{Arnaud}(1996)}]{Arnaud1996}
{Arnaud}, K.~A. 1996, in Astronomical Society of the Pacific Conference Series,
  Vol. 101, Astronomical Data Analysis Software and Systems V, ed. G.~H.
  {Jacoby} \& J.~{Barnes}, 17

\bibitem[{{Atteia}(2022)}]{Atteia2022}
{Atteia}, J.~L. 2022, GRB Coordinates Network, 31365, 1

\bibitem[{{Band} {et~al.}(1993){Band}, {Matteson}, {Ford}, {Schaefer},
  {Palmer}, {Teegarden}, {Cline}, {Briggs}, {Paciesas}, {Pendleton}, {Fishman},
  {Kouveliotou}, {Meegan}, {Wilson}, \& {Lestrade}}]{Band1993}
{Band}, D., {Matteson}, J., {Ford}, L., {et~al.} 1993, Astrophys.~J., 413, 281

\bibitem[{Barbiellini {et~al.}(2001)}]{Barbiellini2001b}
Barbiellini, G., {et~al.} 2001, in GAMMA 2001: Gamma-Ray Astrophysics 2001, AIP
  Conference Proceedings Volume 587, 754--758

\bibitem[{Blandford \& McKee(1976)}]{Blandford1976}
Blandford, R., \& McKee, C. 1976, The physics of Fluids, 19, 1130

\bibitem[{{Bo{\v{s}}njak} {et~al.}(2009){Bo{\v{s}}njak}, {Daigne}, \&
  {Dubus}}]{Bosnjak2009}
{Bo{\v{s}}njak}, {\v{Z}}., {Daigne}, F., \& {Dubus}, G. 2009, \aap, 498, 677,
  \dodoi{10.1051/0004-6361/200811375}

\bibitem[{{Bulgarelli}(2019{\natexlab{a}})}]{Bulgarelli2019a}
{Bulgarelli}, A. 2019{\natexlab{a}}, Experimental Astronomy, 48, 199,
  \dodoi{10.1007/s10686-019-09644-w}

\bibitem[{{Bulgarelli}(2019{\natexlab{b}})}]{Bulgarelli2019b}
---. 2019{\natexlab{b}}, Rendiconti Lincei. Scienze Fisiche e Naturali, 30,
  207, \dodoi{10.1007/s12210-019-00860-2}

\bibitem[{{Caballero-Garcia} {et~al.}(2022){Caballero-Garcia},
  {Sanchez-Ramirez}, {Hu}, {Castro-Tirado}, {Fernandez-Garcia}, {Bergond}, \&
  {Hermelo}}]{Caballero-Garcia2022}
{Caballero-Garcia}, M.~D., {Sanchez-Ramirez}, R., {Hu}, Y.~D., {et~al.} 2022,
  GRB Coordinates Network, 31388, 1

\bibitem[{{Connaughton} {et~al.}(2017){Connaughton}, {Goldstein}, \& {Fermi GBM
  - LIGO Group}}]{Connaughton2017}
{Connaughton}, V., {Goldstein}, A., \& {Fermi GBM - LIGO Group}. 2017, in
  American Astronomical Society Meeting Abstracts, Vol. 229, American
  Astronomical Society Meeting Abstracts \#229, 406.08

\bibitem[{{D'Ai} {et~al.}(2022){D'Ai}, {Sbarufatti}, {Burrows}, {Gropp},
  {Osborne}, {Page}, {Beardmore}, {Melandri}, {Sbarrato}, {Tohuvavohu}, \&
  {Swift-XRT Team}}]{Dai2022}
{D'Ai}, A., {Sbarufatti}, B., {Burrows}, D.~N., {et~al.} 2022, GRB Coordinates
  Network, 31355, 1

\bibitem[{{De Pasquale} {et~al.}(2010){De Pasquale}, {Schady}, {Kuin}, {Page},
  {Curran}, {Zane}, {Oates}, {Holland}, {Breeveld}, {Hoversten}, {Chincarini},
  {Grupe}, {Abdo}, {Ackermann}, {Ajello}, {Axelsson}, {Baldini}, {Ballet},
  {Barbiellini}, {Baring}, {Bastieri}, {Bechtol}, {Bellazzini}, {Berenji},
  {Bissaldi}, {Blandford}, {Bloom}, {Bonamente}, {Borgland}, {Bouvier},
  {Bregeon}, {Brez}, {Briggs}, {Brigida}, {Bruel}, {Burnett}, {Buson},
  {Caliandro}, {Cameron}, {Caraveo}, {Carrigan}, {Casandjian}, {Cecchi},
  {{\c{C}}elik}, {Chekhtman}, {Chiang}, {Ciprini}, {Claus}, {Cohen-Tanugi},
  {Connaughton}, {Conrad}, {Dermer}, {de Angelis}, {de Palma}, {Dingus},
  {Silva}, {Drell}, {Dubois}, {Dumora}, {Farnier}, {Favuzzi}, {Fegan},
  {Fishman}, {Focke}, {Frailis}, {Fukazawa}, {Funk}, {Fusco}, {Gargano},
  {Gasparrini}, {Gehrels}, {Germani}, {Giglietto}, {Giordano}, {Glanzman},
  {Godfrey}, {Granot}, {Greiner}, {Grenier}, {Grove}, {Guillemot}, {Guiriec},
  {Harding}, {Hayashida}, {Hays}, {Horan}, {Hughes}, {Jackson},
  {J{\'o}hannesson}, {Johnson}, {Johnson}, {Kamae}, {Katagiri}, {Kataoka},
  {Kawai}, {Kerr}, {Kippen}, {Kn{\"o}dlseder}, {Kocevski}, {Kuss}, {Lande},
  {Latronico}, {Lemoine-Goumard}, {Longo}, {Loparco}, {Lott}, {Lovellette},
  {Lubrano}, {Makeev}, {Mazziotta}, {McEnery}, {McGlynn}, {Meegan},
  {M{\'e}sz{\'a}ros}, {Meurer}, {Michelson}, {Mitthumsiri}, {Mizuno}, {Monte},
  {Monzani}, {Moretti}, {Morselli}, {Moskalenko}, {Murgia}, {Nolan}, {Norris},
  {Nuss}, {Ohno}, {Ohsugi}, {Omodei}, {Orlando}, {Ormes}, {Paciesas},
  {Paneque}, {Panetta}, {Parent}, {Pelassa}, {Pepe}, {Pesce-Rollins}, {Piron},
  {Porter}, {Preece}, {Rain{\`o}}, {Rando}, {Razzano}, {Reimer}, {Reimer},
  {Reposeur}, {Ritz}, {Rochester}, {Rodriguez}, {Roth}, {Ryde}, {Sadrozinski},
  {Sander}, {Saz Parkinson}, {Scargle}, {Schalk}, {Sgr{\`o}}, {Siskind},
  {Smith}, {Spandre}, {Spinelli}, {Stamatikos}, {Starck}, {Stecker},
  {Strickman}, {Suson}, {Tajima}, {Takahashi}, {Tanaka}, {Thayer}, {Thayer},
  {Thompson}, {Tibaldo}, {Toma}, {Torres}, {Tosti}, {Tramacere}, {Uchiyama},
  {Uehara}, {Usher}, {van der Horst}, {Vasileiou}, {Vilchez}, {Vitale}, {von
  Kienlin}, {Waite}, {Wang}, {Winer}, {Wood}, {Wu}, {Yamazaki}, {Ylinen}, \&
  {Ziegler}}]{Depasquale2010}
{De Pasquale}, M., {Schady}, P., {Kuin}, N.~P.~M., {et~al.} 2010, \apjl, 709,
  L146, \dodoi{10.1088/2041-8205/709/2/L146}

\bibitem[{{Derishev} \& {Piran}(2021)}]{Derishev&Piran2021}
{Derishev}, E., \& {Piran}, T. 2021, \apj, 923, 135,
  \dodoi{10.3847/1538-4357/ac2dec}

\bibitem[{{Dimple} {et~al.}(2022){Dimple}, {Ghosh}, {Gupta}, {Kumar}, {Ror},
  {Panchal}, {Misra}, \& {Pandey}}]{Dimple2022}
{Dimple}, {Ghosh}, A., {Gupta}, R., {et~al.} 2022, GRB Coordinates Network,
  31384, 1

\bibitem[{{Fu} {et~al.}(2022){Fu}, {Zhu}, {Xu}, {Liu}, \& {Jiang}}]{Fu2022}
{Fu}, S.~Y., {Zhu}, Z.~P., {Xu}, D., {Liu}, X., \& {Jiang}, S.~Q. 2022, GRB
  Coordinates Network, 31353, 1

\bibitem[{{Fynbo} {et~al.}(2022){Fynbo}, {de Ugarte Postigo}, {Xu}, {Malesani},
  {Milvang-Jensen}, \& {Viuho}}]{Fynbo2022}
{Fynbo}, J.~P.~U., {de Ugarte Postigo}, A., {Xu}, D., {et~al.} 2022, GRB
  Coordinates Network, 31359, 1

\bibitem[{{Galli} {et~al.}(2013){Galli}, {Marisaldi}, {Fuschino}, {Labanti},
  {Argan}, {Barbiellini}, {Bulgarelli}, {Cattaneo}, {Colafrancesco}, {Del
  Monte}, {Feroci}, {Gianotti}, {Giuliani}, {Longo}, {Mereghetti}, {Morselli},
  {Pacciani}, {Pellizzoni}, {Pittori}, {Rapisarda}, {Rappoldi}, {Tavani},
  {Trifoglio}, {Trois}, {Vercellone}, \& {Verrecchia}}]{Galli2013}
{Galli}, M., {Marisaldi}, M., {Fuschino}, F., {et~al.} 2013,
  Astron.~Astrophys., 553, A33, \dodoi{10.1051/0004-6361/201220833}

\bibitem[{{Gehrels} \& {M{\'e}sz{\'a}ros}(2012)}]{Gehrels&Meszaros2012}
{Gehrels}, N., \& {M{\'e}sz{\'a}ros}, P. 2012, Science, 337, 932,
  \dodoi{10.1126/science.1216793}

\bibitem[{{Giuliani} {et~al.}(2008){Giuliani}, {Mereghetti}, {Fornari}, {Del
  Monte}, {Feroci}, {Marisaldi}, {Esposito}, {Perotti}, {Tavani}, {Argan},
  {Barbiellini}, {Boffelli}, {Bulgarelli}, {Caraveo}, {Cattaneo}, {Chen},
  {Costa}, {D'Ammando}, {Di Cocco}, {Donnarumma}, {Evangelista}, {Fiorini},
  {Fuschino}, {Galli}, {Gianotti}, {Labanti}, {Lapshov}, {Lazzarotto},
  {Lipari}, {Longo}, {Morselli}, {Pacciani}, {Pellizzoni}, {Piano}, {Picozza},
  {Prest}, {Pucella}, {Rapisarda}, {Rappoldi}, {Soffitta}, {Trifoglio},
  {Trois}, {Vallazza}, {Vercellone}, {Zanello}, {Salotti}, {Cutini}, {Pittori},
  {Preger}, {Santolamazza}, {Verrecchia}, {Gehrels}, {Page}, {Burrows},
  {Rossi}, {Hurley}, {Mitrofanov}, \& {Boynton}}]{Giuliani2008}
{Giuliani}, A., {Mereghetti}, S., {Fornari}, F., {et~al.} 2008, \aap, 491, L25,
  \dodoi{10.1051/0004-6361:200810737}

\bibitem[{{Giuliani} {et~al.}(2010){Giuliani}, {Fuschino}, {Vianello},
  {Marisaldi}, {Mereghetti}, {Tavani}, {Cutini}, {Barbiellini}, {Longo},
  {Moretti}, {Feroci}, {Del Monte}, {Argan}, {Bulgarelli}, {Caraveo},
  {Cattaneo}, {Chen}, {Contessi}, {D'Ammand o}, {Costa}, {De Paris}, {Di
  Cocco}, {Donnarumma}, {Evangelista}, {Ferrari}, {Fiorini}, {Galli},
  {Gianotti}, {Labanti}, {Lapshov}, {Lazzarotto}, {Lipari}, {Morselli},
  {Pacciani}, {Pellizzoni}, {Perotti}, {Piano}, {Picozza}, {Pilia}, {Pucella},
  {Prest}, {Rapisarda}, {Rappoldi}, {Rubini}, {Sabatini}, {Scalise}, {Striani},
  {Soffitta}, {Trifoglio}, {Trois}, {Vallazza}, {Vercellone}, {Vittorini},
  {Zambra}, {Zanello}, {Pittori}, {Verrecchia}, {Santolamazza}, {Giommi},
  {Colafrancesco}, {Antonelli}, \& {Salotti}}]{Giuliani2010}
{Giuliani}, A., {Fuschino}, F., {Vianello}, G., {et~al.} 2010, \apjl, 708, L84,
  \dodoi{10.1088/2041-8205/708/2/L84}

\bibitem[{{Giuliani} {et~al.}(2014){Giuliani}, {Mereghetti}, {Marisaldi},
  {Longo}, {Del Monte}, {Pittori}, {Verrecchia}, {Tavani}, {Cattaneo},
  {Pacciani}, {Vercellone}, \& {Rappoldi}}]{Giuliani2014}
{Giuliani}, A., {Mereghetti}, S., {Marisaldi}, M., {et~al.} 2014, arXiv
  e-prints, arXiv:1407.0238.
\newblock \doarXiv{1407.0238}

\bibitem[{{Goldstein} {et~al.}(2017){Goldstein}, {Veres}, {Burns}, {Briggs},
  {Hamburg}, {Kocevski}, {Wilson-Hodge}, {Preece}, {Poolakkil}, {Roberts},
  {Hui}, {Connaughton}, {Racusin}, {von Kienlin}, {Dal Canton}, {Christensen},
  {Littenberg}, {Siellez}, {Blackburn}, {Broida}, {Bissaldi}, {Cleveland},
  {Gibby}, {Giles}, {Kippen}, {McBreen}, {McEnery}, {Meegan}, {Paciesas}, \&
  {Stanbro}}]{Goldstein2017}
{Goldstein}, A., {Veres}, P., {Burns}, E., {et~al.} 2017, \apjl, 848, L14,
  \dodoi{10.3847/2041-8213/aa8f41}

\bibitem[{{Gonz{\'a}lez} {et~al.}(2003){Gonz{\'a}lez}, {Dingus}, {Kaneko},
  {Preece}, {Dermer}, \& {Briggs}}]{Gonzalez2003}
{Gonz{\'a}lez}, M.~M., {Dingus}, B.~L., {Kaneko}, Y., {et~al.} 2003, \nat, 424,
  749, \dodoi{10.1038/nature01869}

\bibitem[{{Klebesadel} {et~al.}(1973){Klebesadel}, {Strong}, \&
  {Olson}}]{Klebesadel1973}
{Klebesadel}, R.~W., {Strong}, I.~B., \& {Olson}, R.~A. 1973, Astrophys.~J.l,
  182, L85, \dodoi{10.1086/181225}

\bibitem[{{Koshut} {et~al.}(1996){Koshut}, {Paciesas}, {Kouveliotou}, {van
  Paradijs}, {Pendleton}, {Fishman}, \& {Meegan}}]{Koshut1996}
{Koshut}, T.~M., {Paciesas}, W.~S., {Kouveliotou}, C., {et~al.} 1996, \apj,
  463, 570, \dodoi{10.1086/177272}

\bibitem[{{Kouveliotou} {et~al.}(1993){Kouveliotou}, {Meegan}, {Fishman},
  {Bhat}, {Briggs}, {Koshut}, {Paciesas}, \& {Pendleton}}]{Kouveliotou1993}
{Kouveliotou}, C., {Meegan}, C.~A., {Fishman}, G.~J., {et~al.} 1993,
  Astrophys.~J.l, 413, L101, \dodoi{10.1086/186969}

\bibitem[{{Kuin} {et~al.}(2022){Kuin}, {Tohuvavohu}, \& {Swift/UVOT
  Team}}]{Kuin2022}
{Kuin}, N.~P.~M., {Tohuvavohu}, A., \& {Swift/UVOT Team}. 2022, GRB Coordinates
  Network, 31351, 1

\bibitem[{{Labanti} {et~al.}(2009){Labanti}, {Marisaldi}, {Fuschino}, {Galli},
  {Argan}, {Bulgarelli}, {di Cocco}, {Gianotti}, {Tavani}, \&
  {Trifoglio}}]{Labanti2009}
{Labanti}, C., {Marisaldi}, M., {Fuschino}, F., {et~al.} 2009, Nuclear
  Instruments and Methods in Physics Research A, 598, 470,
  \dodoi{10.1016/j.nima.2008.09.021}

\bibitem[{{Lesage} {et~al.}(2022){Lesage}, {Meegan}, \& {Fermi Gamma-ray Burst
  Monitor Team}}]{Lesage2022}
{Lesage}, S., {Meegan}, C., \& {Fermi Gamma-ray Burst Monitor Team}. 2022, GRB
  Coordinates Network, 31360, 1

\bibitem[{{MAGIC Collaboration} {et~al.}(2019){MAGIC Collaboration}, {Acciari},
  {Ansoldi}, {Antonelli}, {Arbet Engels}, {Baack}, {Babi{\'c}}, {Banerjee},
  {Barres de Almeida}, {Barrio}, {Becerra Gonz{\'a}lez}, {Bednarek},
  {Bellizzi}, {Bernardini}, {Berti}, {Besenrieder}, {Bhattacharyya},
  {Bigongiari}, {Biland}, {Blanch}, {Bonnoli}, {Bo{\v{s}}njak}, {Busetto},
  {Carosi}, {Carosi}, {Ceribella}, {Chai}, {Chilingaryan}, {Cikota}, {Colak},
  {Colin}, {Colombo}, {Contreras}, {Cortina}, {Covino}, {D'Amico}, {D'Elia},
  {da Vela}, {Dazzi}, {de Angelis}, {de Lotto}, {Delfino}, {Delgado},
  {Depaoli}, {di Pierro}, {di Venere}, {Do Souto Espi{\~n}eira}, {Dominis
  Prester}, {Donini}, {Dorner}, {Doro}, {Elsaesser}, {Fallah Ramazani},
  {Fattorini}, {Fern{\'a}ndez-Barral}, {Ferrara}, {Fidalgo}, {Foffano},
  {Fonseca}, {Font}, {Fruck}, {Fukami}, {Gallozzi}, {Garc{\'\i}a L{\'o}pez},
  {Garczarczyk}, {Gasparyan}, {Gaug}, {Giglietto}, {Giordano}, {Godinovi{\'c}},
  {Green}, {Guberman}, {Hadasch}, {Hahn}, {Herrera}, {Hoang}, {Hrupec},
  {H{\"u}tten}, {Inada}, {Inoue}, {Ishio}, {Iwamura}, {Jouvin}, {Kerszberg},
  {Kubo}, {Kushida}, {Lamastra}, {Lelas}, {Leone}, {Lindfors}, {Lombardi},
  {Longo}, {L{\'o}pez}, {L{\'o}pez-Coto}, {L{\'o}pez-Oramas}, {Loporchio},
  {Machado de Oliveira Fraga}, {Maggio}, {Majumdar}, {Makariev}, {Mallamaci},
  {Maneva}, {Manganaro}, {Mannheim}, {Maraschi}, {Mariotti}, {Mart{\'\i}nez},
  {Masuda}, {Mazin}, {Mi{\'c}anovi{\'c}}, {Miceli}, {Minev}, {Miranda},
  {Mirzoyan}, {Molina}, {Moralejo}, {Morcuende}, {Moreno}, {Moretti},
  {Munar-Adrover}, {Neustroev}, {Nigro}, {Nilsson}, {Ninci}, {Nishijima},
  {Noda}, {Nogu{\'e}s}, {N{\"o}the}, {Nozaki}, {Paiano}, {Palacio},
  {Palatiello}, {Paneque}, {Paoletti}, {Paredes}, {Pe{\~n}il}, {Peresano},
  {Persic}, {Prada Moroni}, {Prandini}, {Puljak}, {Rhode}, {Rib{\'o}}, {Rico},
  {Righi}, {Rugliancich}, {Saha}, {Sahakyan}, {Saito}, {Sakurai}, {Satalecka},
  {Schmidt}, {Schweizer}, {Sitarek}, {{\v{S}}nidari{\'c}}, {Sobczynska},
  {Somero}, {Stamerra}, {Strom}, {Strzys}, {Suda}, {Suri{\'c}}, {Takahashi},
  {Tavecchio}, {Temnikov}, {Terzi{\'c}}, {Teshima}, {Torres-Alb{\`a}}, {Tosti},
  {Tsujimoto}, {Vagelli}, {van Scherpenberg}, {Vanzo}, {Vazquez Acosta},
  {Vigorito}, {Vitale}, {Vovk}, {Will}, {Zari{\'c}}, \& {Nava}}]{MAGIC2019}
{MAGIC Collaboration}, {Acciari}, V.~A., {Ansoldi}, S., {et~al.} 2019, \nat,
  575, 455, \dodoi{10.1038/s41586-019-1750-x}

\bibitem[{Maiorana {et~al.}(2020{\natexlab{a}})Maiorana, Marisaldi, Lindanger,
  Østgaard, Ursi, Sarria, Galli, Labanti, Tavani, Pittori, \&
  Verrecchia}]{Maiorana2020}
Maiorana, C., Marisaldi, M., Lindanger, A., {et~al.} 2020{\natexlab{a}},
  Journal of Geophysical Research: Atmospheres, 125, e2019JD031986,
  \dodoi{https://doi.org/10.1029/2019JD031986}

\bibitem[{Maiorana {et~al.}(2020{\natexlab{b}})Maiorana, Marisaldi, Lindanger,
  Østgaard, Ursi, Sarria, Galli, Labanti, Tavani, Pittori, \&
  Verrecchia}]{Lindanger2020}
---. 2020{\natexlab{b}}, Journal of Geophysical Research: Atmospheres, 125,
  e2019JD031986, \dodoi{https://doi.org/10.1029/2019JD031986}

\bibitem[{Marisaldi {et~al.}(2008)Marisaldi, Labanti, Fuschino,
  {et~al.}}]{Marisaldi2008}
Marisaldi, M., Labanti, C., Fuschino, F., {et~al.} 2008, proc. of Gamma Ray
  Bursts 2007, November 5-9, Santa Fe, NM, AIP Conf. Proc., 1000, 531

\bibitem[{{Marisaldi} {et~al.}(2014){Marisaldi}, {Fuschino}, {Tavani},
  {Dietrich}, {Price}, {Galli}, {Pittori}, {Verrecchia}, {Mereghetti},
  {Cattaneo}, {Colafrancesco}, {Argan}, {Labanti}, {Longo}, {Del Monte},
  {Barbiellini}, {Giuliani}, {Bulgarelli}, {Campana}, {Chen}, {Gianotti},
  {Giommi}, {Lazzarotto}, {Morselli}, {Rapisarda}, {Rappoldi}, {Trifoglio},
  {Trois}, \& {Vercellone}}]{Marisaldi2014}
{Marisaldi}, M., {Fuschino}, F., {Tavani}, M., {et~al.} 2014, Journal of
  Geophysical Research (Space Physics), 119, 1337, \dodoi{10.1002/2013JA019301}

\bibitem[{{Markwardt} {et~al.}(2022){Markwardt}, {Barthelmy}, {Krimm}, {Laha},
  {Lien}, {Palmer}, {Parsotan}, {Sakamoto}, \& {Stamatikos}}]{Markwardt2022}
{Markwardt}, C.~B., {Barthelmy}, S.~D., {Krimm}, H.~A., {et~al.} 2022, GRB
  Coordinates Network, 31369, 1

\bibitem[{{Moskvitin} {et~al.}(2022){Moskvitin}, {Pankov}, {Medvedev},
  {Belkin}, {Pozanenko}, \& {GRB IKI FuN}}]{Moskvitin2022}
{Moskvitin}, A.~S., {Pankov}, N., {Medvedev}, A.~S., {et~al.} 2022, GRB
  Coordinates Network, 31420, 1

\bibitem[{{Nicuesa Guelbenzu} {et~al.}(2022{\natexlab{a}}){Nicuesa Guelbenzu},
  {Klose}, {Melnikov}, {Stecklum}, \& {Laux}}]{NicuesaGuelbenzu2022a}
{Nicuesa Guelbenzu}, A., {Klose}, S., {Melnikov}, S., {Stecklum}, B., \&
  {Laux}, U. 2022{\natexlab{a}}, GRB Coordinates Network, 31370, 1

\bibitem[{{Nicuesa Guelbenzu} {et~al.}(2022{\natexlab{b}}){Nicuesa Guelbenzu},
  {Melnikov}, {Klose}, {Stecklum}, \& {Ludwig}}]{NicuesaGuelbenzu2022b}
{Nicuesa Guelbenzu}, A., {Melnikov}, S., {Klose}, S., {Stecklum}, B., \&
  {Ludwig}, F. 2022{\natexlab{b}}, GRB Coordinates Network, 31401, 1

\bibitem[{{Osborne} {et~al.}(2022){Osborne}, {Beardmore}, {Evans}, {Goad}, \&
  {Swift-XRT Team.}}]{Osborne2022}
{Osborne}, J.~P., {Beardmore}, A.~P., {Evans}, P.~A., {Goad}, M.~R., \&
  {Swift-XRT Team.} 2022, GRB Coordinates Network, 31349, 1

\bibitem[{Parmiggiani {et~al.}(2021)Parmiggiani, Bulgarelli, Ursi, Fioretti,
  Baroncelli, Addis, Di~Piano, Pittori, Verrecchia, Lucarelli, Tavani, \&
  Beneventano}]{Parmiggiani2021}
Parmiggiani, N., Bulgarelli, A., Ursi, A., {et~al.} 2021, PoS, ICRC2021, 933,
  \dodoi{10.22323/1.395.0933}

\bibitem[{{Perley}(2022)}]{Perley2022}
{Perley}, D.~A. 2022, GRB Coordinates Network, 31357, 1

\bibitem[{{Piran}(1999)}]{Piran1999}
{Piran}, T. 1999, \physrep, 314, 575, \dodoi{10.1016/S0370-1573(98)00127-6}

\bibitem[{{Piron}(2016)}]{Piron2016}
{Piron}, F. 2016, Comptes Rendus Physique, 17, 617,
  \dodoi{10.1016/j.crhy.2016.04.005}

\bibitem[{{Pittori} \& {The Agile-Ssdc Team}(2019)}]{Pittori2019}
{Pittori}, C., \& {The Agile-Ssdc Team}. 2019, Rendiconti Lincei. Scienze
  Fisiche e Naturali, 30, 217, \dodoi{10.1007/s12210-019-00857-x}

\bibitem[{{Planck Collaboration} {et~al.}(2014){Planck Collaboration}, {Ade},
  {Arnaud}, {Ashdown}, {Aumont}, {Baccigalupi}, {Banday}, {Barreiro},
  {Battaner}, {Benabed}, {Benoit-L{\'e}vy}, {Bernard}, {Bersanelli},
  {Bielewicz}, {Bond}, {Borrill}, {Bouchet}, {Burigana}, {Cardoso}, {Catalano},
  {Challinor}, {Chamballu}, {Chiang}, {Christensen}, {Clements}, {Colombi},
  {Colombo}, {Couchot}, {Coulais}, {Crill}, {Curto}, {Cuttaia}, {Danese},
  {Davies}, {Davis}, {de Bernardis}, {de Rosa}, {de Zotti}, {Delabrouille},
  {D{\'e}sert}, {Dickinson}, {Diego}, {Dole}, {Donzelli}, {Dor{\'e}},
  {Douspis}, {Dupac}, {En{\ss}lin}, {Eriksen}, {Finelli}, {Forni}, {Frailis},
  {Fraisse}, {Franceschi}, {Galeotta}, {Ganga}, {Giard}, {Gonz{\'a}lez-Nuevo},
  {G{\'o}rski}, {Gratton}, {Gregorio}, {Gruppuso}, {Gudmundsson}, {Hansen},
  {Hanson}, {Harrison}, {Henrot-Versill{\'e}}, {Herranz}, {Hildebrandt},
  {Hivon}, {Hobson}, {Holmes}, {Hornstrup}, {Hovest}, {Huffenberger}, {Jaffe},
  {Jaffe}, {Jones}, {Keih{\"a}nen}, {Keskitalo}, {Knoche}, {Kunz},
  {Kurki-Suonio}, {Lagache}, {L{\"a}hteenm{\"a}ki}, {Lamarre}, {Lasenby},
  {Lawrence}, {Leonardi}, {Le{\'o}n-Tavares}, {Lesgourgues}, {Liguori},
  {Lilje}, {Linden-V{\o}rnle}, {L{\'o}pez-Caniego}, {Lubin},
  {Mac{\'\i}as-P{\'e}rez}, {Maino}, {Mandolesi}, {Maris}, {Martin},
  {Mart{\'\i}nez-Gonz{\'a}lez}, {Masi}, {Matarrese}, {Mazzotta}, {Meinhold},
  {Melchiorri}, {Mendes}, {Mennella}, {Migliaccio}, {Mitra},
  {Miville-Desch{\^e}nes}, {Moneti}, {Montier}, {Morgante}, {Mortlock}, {Moss},
  {Munshi}, {Murphy}, {Naselsky}, {Nati}, {Natoli}, {N{\o}rgaard-Nielsen},
  {Noviello}, {Novikov}, {Novikov}, {Oxborrow}, {Pagano}, {Pajot}, {Paoletti},
  {Partridge}, {Pasian}, {Patanchon}, {Pearson}, {Pearson}, {Perdereau},
  {Perrotta}, {Piacentini}, {Piat}, {Pierpaoli}, {Pietrobon}, {Plaszczynski},
  {Pointecouteau}, {Polenta}, {Ponthieu}, {Popa}, {Pratt}, {Prunet}, {Puget},
  {Rachen}, {Reinecke}, {Remazeilles}, {Renault}, {Ricciardi}, {Ristorcelli},
  {Rocha}, {Roudier}, {Rubi{\~n}o-Mart{\'\i}n}, {Rusholme}, {Sandri}, {Scott},
  {Stolyarov}, {Sudiwala}, {Sutton}, {Suur-Uski}, {Sygnet}, {Tauber},
  {Terenzi}, {Toffolatti}, {Tomasi}, {Tristram}, {Tucci}, {Valenziano},
  {Valiviita}, {Van Tent}, {Vielva}, {Villa}, {Wade}, {Wandelt}, {Wehus},
  {White}, {Yvon}, {Zacchei}, \& {Zonca}}]{Ade2014}
{Planck Collaboration}, {Ade}, P.~A.~R., {Arnaud}, M., {et~al.} 2014, \aap,
  571, A31, \dodoi{10.1051/0004-6361/201423743}

\bibitem[{{Prest} {et~al.}(2003){Prest}, {Barbiellini}, {Bordignon}, {Fedel},
  {Liello}, {Longo}, {Pontoni}, \& {Vallazza}}]{Prest2003}
{Prest}, M., {Barbiellini}, G., {Bordignon}, G., {et~al.} 2003, Nuclear
  Instruments and Methods in Physics Research A, 501, 280,
  \dodoi{10.1016/S0168-9002(02)02047-8}

\bibitem[{{Ror} {et~al.}(2022){Ror}, {Gupta}, {Kumar}, {Dimple}, {Ghosh},
  {Aryan}, {Kumar}, {Pandey}, \& {Misra}}]{Ror2022}
{Ror}, A., {Gupta}, R., {Kumar}, A., {et~al.} 2022, GRB Coordinates Network,
  31394, 1

\bibitem[{{Ruffini} {et~al.}(2022{\natexlab{a}}){Ruffini}, {Aimuratov},
  {Becerra}, {Bianco}, {Chen}, {Cherubini}, {Cai}, {Eslamzadeh}, {Filippi},
  {Karlica}, {Li}, {Mathews}, {Moradi}, {Muccino}, {Pisani}, {Rastegar Nia},
  {Rueda}, {Sahakyan}, {Wang}, {Xue}, {Yuan}, {Zheng}, {Icra}, {Icranet}, \&
  {Ustc Team}}]{Ruffini2022b}
{Ruffini}, R., {Aimuratov}, Y., {Becerra}, L., {et~al.} 2022{\natexlab{a}}, GRB
  Coordinates Network, 31648, 1

\bibitem[{{Ruffini} {et~al.}(2022{\natexlab{b}}){Ruffini}, {Aimuratov},
  {Becerra}, {Bianco}, {Chen}, {Cherubini}, {Cai}, {Eslamzadeh}, {Filippi},
  {Karlica}, {Li}, {Mathews}, {Moradi}, {Muccino}, {Pisani}, {Rastegar Nia},
  {Rueda}, {Sahakyan}, {Wang}, {Xue}, {Yuan}, {Zheng}, {Icra}, {Icranet}, \&
  {Ustc Team}}]{Ruffini2022a}
---. 2022{\natexlab{b}}, GRB Coordinates Network, 31465, 1

\bibitem[{{Schneid} {et~al.}(1992){Schneid}, {Bertsch}, {Fichtel}, {Hartman},
  {Hunter}, {Kanbach}, {Kniffen}, {Kwok}, {Lin}, {Mattox},
  {Mayer-Hasselwander}, {Michelson}, {von Montigny}, {Nolan}, {Pinkau},
  {Rothermel}, {Sommer}, {Sreekumar}, \& {Thompson}}]{Schneid1992}
{Schneid}, E.~J., {Bertsch}, D.~L., {Fichtel}, C.~E., {et~al.} 1992, \aap, 255,
  L13

\bibitem[{{Sommer} {et~al.}(1994){Sommer}, {Bertsch}, {Dingus}, {Fichtel},
  {Fishman}, {Harding}, {Hartman}, {Hunter}, {Hurley}, {Kanbach}, {Kniffen},
  {Kouveliotou}, {Lin}, {Mattox}, {Mayer-Hasselwander}, {Michelson}, {von
  Montigny}, {Nolan}, {Schneid}, {Sreekumar}, \& {Thompson}}]{Sommer1994}
{Sommer}, M., {Bertsch}, D.~L., {Dingus}, B.~L., {et~al.} 1994, \apjl, 422,
  L63, \dodoi{10.1086/187213}

\bibitem[{{Strausbaugh} \& {Cucchiara}(2022{\natexlab{a}})}]{Strausbaugh2022a}
{Strausbaugh}, R., \& {Cucchiara}, A. 2022{\natexlab{a}}, GRB Coordinates
  Network, 31371, 1

\bibitem[{{Strausbaugh} \& {Cucchiara}(2022{\natexlab{b}})}]{Strausbaugh2022b}
---. 2022{\natexlab{b}}, GRB Coordinates Network, 31383, 1

\bibitem[{{Strausbaugh} \& {Cucchiara}(2022{\natexlab{c}})}]{Strausbaugh2022c}
---. 2022{\natexlab{c}}, GRB Coordinates Network, 31389, 1

\bibitem[{{Tavani} {et~al.}(2009){Tavani}, {Barbiellini}, {Argan}, {Boffelli},
  {Bulgarelli}, {Caraveo}, {Cattaneo}, {Chen}, {Cocco}, {Costa}, {D'Ammando},
  {Del Monte}, {de Paris}, {Di Cocco}, {di Persio}, {Donnarumma},
  {Evangelista}, {Feroci}, {Ferrari}, {Fiorini}, {Fornari}, {Fuschino},
  {Froysland}, {Frutti}, {Galli}, {Gianotti}, {Giuliani}, {Labanti}, {Lapshov},
  {Lazzarotto}, {Liello}, {Lipari}, {Longo}, {Mattaini}, {Marisaldi},
  {Mastropietro}, {Mauri}, {Mauri}, {Mereghetti}, {Morelli}, {Morselli},
  {Pacciani}, {Pellizzoni}, {Perotti}, {Piano}, {Picozza}, {Pontoni},
  {Porrovecchio}, {Prest}, {Pucella}, {Rapisarda}, {Rappoldi}, {Rossi},
  {Rubini}, {Soffitta}, {Traci}, {Trifoglio}, {Trois}, {Vallazza},
  {Vercellone}, {Vittorini}, {Zambra}, {Zanello}, {Pittori}, {Preger},
  {Santolamazza}, {Verrecchia}, {Giommi}, {Colafrancesco}, {Antonelli},
  {Cutini}, {Gasparrini}, {Stellato}, {Fanari}, {Primavera}, {Tamburelli},
  {Viola}, {Guarrera}, {Salotti}, {D'Amico}, {Marchetti}, {Crisconio},
  {Sabatini}, {Annoni}, {Alia}, {Longoni}, {Sanquerin}, {Battilana}, {Concari},
  {Dessimone}, {Grossi}, {Parise}, {Monzani}, {Artina}, {Pavesi},
  {Marseguerra}, {Nicolini}, {Scandelli}, {Soli}, {Vettorello}, {Zardetto},
  {Bonati}, {Maltecca}, {D'Alba}, {Patan{\'e}}, {Babini}, {Onorati},
  {Acquaroli}, {Angelucci}, {Morelli}, {Agostara}, {Cerone}, {Michetti},
  {Tempesta}, {D'Eramo}, {Rocca}, {Giannini}, {Borghi}, {Garavelli}, {Conte},
  {Balasini}, {Ferrario}, {Vanotti}, {Collavo}, \& {Giacomazzo}}]{Tavani2008b}
{Tavani}, M., {Barbiellini}, G., {Argan}, A., {et~al.} 2009,
  Astron.~Astrophys., 502, 995, \dodoi{10.1051/0004-6361/200810527}

\bibitem[{{Tavani} {et~al.}(2021){Tavani}, {Casentini}, {Ursi}, {Verrecchia},
  {Addis}, {Antonelli}, {Argan}, {Barbiellini}, {Baroncelli}, {Bernardi},
  {Bianchi}, {Bulgarelli}, {Caraveo}, {Cardillo}, {Cattaneo}, {Chen}, {Costa},
  {Del Monte}, {Di Cocco}, {Di Persio}, {Donnarumma}, {Evangelista}, {Feroci},
  {Ferrari}, {Fioretti}, {Fuschino}, {Galli}, {Gianotti}, {Giuliani},
  {Labanti}, {Lazzarotto}, {Lipari}, {Longo}, {Lucarelli}, {Magro},
  {Marisaldi}, {Mereghetti}, {Morelli}, {Morselli}, {Naldi}, {Pacciani},
  {Parmiggiani}, {Paoletti}, {Pellizzoni}, {Perri}, {Perotti}, {Piano},
  {Picozza}, {Pilia}, {Pittori}, {Puccetti}, {Pupillo}, {Rapisarda},
  {Rappoldi}, {Rubini}, {Setti}, {Soffitta}, {Trifoglio}, {Trois},
  {Vercellone}, {Vittorini}, {Giommi}, \& {D'Amico}}]{Tavani2021}
{Tavani}, M., {Casentini}, C., {Ursi}, A., {et~al.} 2021, Nature Astronomy, 5,
  401, \dodoi{10.1038/s41550-020-01276-x}

\bibitem[{{Tohuvavohu} {et~al.}(2022){Tohuvavohu}, {Gropp}, {Kennea}, {Lien},
  {Palmer}, {Parsotan}, {Sbarufatti}, {Siegel}, \& {Neil Gehrels Swift
  Observatory Team}}]{Tohuvavohu2022}
{Tohuvavohu}, A., {Gropp}, J.~D., {Kennea}, J.~A., {et~al.} 2022, GRB
  Coordinates Network, 31347, 1

\bibitem[{{Tomasella} {et~al.}(2022){Tomasella}, {Brocato}, {D'Onofrio},
  {Cappellaro}, \& {Benetti}}]{Tomasella2022}
{Tomasella}, L., {Brocato}, E., {D'Onofrio}, M., {Cappellaro}, E., \&
  {Benetti}, S. 2022, GRB Coordinates Network, 31363, 1

\bibitem[{{Tsvetkova} {et~al.}(2022){Tsvetkova}, {Frederiks}, {Lysenko},
  {Ridnaia}, {Svinkin}, {Ulanov}, {Cline}, \& {Konus-Wind
  Team}}]{Tsvetkova2022a}
{Tsvetkova}, A., {Frederiks}, D., {Lysenko}, A., {et~al.} 2022, GRB Coordinates
  Network, 31433, 1

\bibitem[{{Tsvetkova} \& {Konus-Wind Team}(2022)}]{Tsvetkova2022b}
{Tsvetkova}, A., \& {Konus-Wind Team}. 2022, GRB Coordinates Network, 31436, 1

\bibitem[{{Ursi} {et~al.}(2019){Ursi}, {Tavani}, {Verrecchia}, {Marisaldi},
  {Argan}, {Trois}, \& {Tempesta}}]{Ursi2019}
{Ursi}, A., {Tavani}, M., {Verrecchia}, F., {et~al.} 2019, \apj, 871, 27,
  \dodoi{10.3847/1538-4357/aaf28f}

\bibitem[{Ursi {et~al.}(2020{\natexlab{a}})Ursi, Tavani, Frederiks, Romani,
  Verrecchia, Marisaldi, Aptekar, Antonelli, Argan, Bulgarelli, Barbiellini,
  Caraveo, Cardillo, Casentini, Cattaneo, Chen, Costa, Donnarumma, Evangelista,
  Feroci, Ferrari, Fuschino, Galli, Giuliani, Labanti, Lazzarotto, Longo,
  Lucarelli, Morselli, Paoletti, Parmiggiani, Piano, Pilia, Pittori, Svinkin,
  Trois, Tsvetkova, Vercellone, \& Vittorini}]{Ursi2020solar}
Ursi, A., Tavani, M., Frederiks, D.~D., {et~al.} 2020{\natexlab{a}}, The
  Astronomer's Telegram, 14236, 1.
\newblock \url{https://ui.adsabs.harvard.edu/abs/2020ATel14236....1U}

\bibitem[{Ursi {et~al.}(2020{\natexlab{b}})Ursi, Tavani, Frederiks, Romani,
  Verrecchia, Marisaldi, Aptekar, Antonelli, Argan, Bulgarelli, Barbiellini,
  Caraveo, Cardillo, Casentini, Cattaneo, Chen, Costa, Donnarumma, Evangelista,
  Feroci, Ferrari, Fuschino, Galli, Giuliani, Labanti, Lazzarotto, Longo,
  Lucarelli, Morselli, Paoletti, Parmiggiani, Piano, Pilia, Pittori, Svinkin,
  Trois, Tsvetkova, Vercellone, \& Vittorini}]{Ursi2020magic}
---. 2020{\natexlab{b}}, The Astrophysical Journal, 904, 133,
  \dodoi{10.3847/1538-4357/abc2d4}

\bibitem[{{Ursi} {et~al.}(2022{\natexlab{a}}){Ursi}, {Menegoni}, {Longo},
  {Pittori}, {Verrecchia}, {Tempesta}, {Tavani}, {Argan}, {Cardillo},
  {Casentini}, {Evangelista}, {Foffano}, {Piano}, {Lucarelli}, {Bulgarelli},
  {di Piano}, {Fioretti}, {Fuschino}, {Parmiggiani}, {Marisaldi}, {Pilia},
  {Trois}, {Donnarumma}, {Giuliani}, \& {Agile Team}}]{Ursi2022gcn}
{Ursi}, A., {Menegoni}, E., {Longo}, F., {et~al.} 2022{\natexlab{a}}, GRB
  Coordinates Network, 31354, 1

\bibitem[{{Ursi} {et~al.}(2022{\natexlab{b}}){Ursi}, {Romani}, {Verrecchia},
  {Pittori}, {Tavani}, {Marisaldi}, {Galli}, {Labanti}, {Parmiggiani},
  {Bulgarelli}, {Addis}, {Baroncelli}, {Cardillo}, {Casentini}, {Cattaneo},
  {Chen}, {Di Piano}, {Fuschino}, {Longo}, {Lucarelli}, {Morselli}, {Piano}, \&
  {Vercellone}}]{Ursi2022catalog}
{Ursi}, A., {Romani}, M., {Verrecchia}, F., {et~al.} 2022{\natexlab{b}}, \apj,
  925, 152, \dodoi{10.3847/1538-4357/ac3df7}

\bibitem[{{Ursi} {et~al.}(2022{\natexlab{c}}){Ursi}, {Verrecchia}, {Piano},
  {Casentini}, {Tavani}, {Bulgarelli}, {Cardillo}, {Longo}, {Lucarelli},
  {Morselli}, {Parmiggiani}, {Pilia}, {Pittori}, \& {Rappoldi}}]{Ursi2022gw}
{Ursi}, A., {Verrecchia}, F., {Piano}, G., {et~al.} 2022{\natexlab{c}}, \apj,
  924, 80, \dodoi{10.3847/1538-4357/ac332f}

\bibitem[{{Vedrenne}(2009)}]{Vedrenne2009}
{Vedrenne}, Gilbert, A.~J.-L. 2009, {Gamma-Ray Bursts: The brightest explosions
  in the Universe}

\bibitem[{{Verrecchia} {et~al.}(2017){Verrecchia}, {Tavani}, {Ursi}, {Argan},
  {Pittori}, {Donnarumma}, {Bulgarelli}, {Fuschino}, {Labanti}, {Marisaldi},
  {Evangelista}, {Minervini}, {Giuliani}, {Cardillo}, {Longo}, {Lucarelli},
  {Munar-Adrover}, {Piano}, {Pilia}, {Fioretti}, {Parmiggiani}, {Trois}, {Del
  Monte}, {Antonelli}, {Barbiellini}, {Caraveo}, {Cattaneo}, {Colafrancesco},
  {Costa}, {D'Amico}, {Feroci}, {Ferrari}, {Morselli}, {Pacciani}, {Paoletti},
  {Pellizzoni}, {Picozza}, {Rappoldi}, \& {Vercellone}}]{Verrecchia2017a}
{Verrecchia}, F., {Tavani}, M., {Ursi}, A., {et~al.} 2017, \apjl, 847, L20,
  \dodoi{10.3847/2041-8213/aa8224}

\end{thebibliography}

\end{document}